%% file: articolo_4.tex
\documentclass[]{aa}
\usepackage{graphicx}
\def\RS{Raymond \& Smith }
\def\mew{Mewe et al paper I to VI}
\def\rlmf{Landini \& Monsignori Fossi }

\def\ra{Mattioli 1988 }
\def\rb{Karim \& Bhalla (1988)}

\def\rd{Chen (1986b) }
\def\re{Chen (1991) }

\def\ro{Romanik (1988) }

\def\sp{{ Solar Phys.} \,}
\def\ad{{ ADNDT}}

\begin{document}

  \thesaurus{20         
              (02.01.3;  
               02.16.1;  
               02.18.6;  
               13.25.3)}  

   \title{Ionization balance for optically thin plasmas: rate coefficients 
for all atoms and ions of the elements H to Ni }


   \author{P. Mazzotta
          \inst{1}, G. Mazzitelli \inst{2}, S. Colafrancesco \inst{3} and
    N. Vittorio \inst{1}}

   \offprints{P. Mazzotta}

   \institute{Dipartimento di Fisica,  Universit\`a di Roma ``Tor Vergata''\\
              via della Ricerca Scientifica 1, I-00133 Roma, Italy
   \and      
              Associazione EURATOM-ENEA sulla Fusione, C.R. Frascati,\\ 
              CP 65-00044 Frascati Roma, Italy
   \and      
              Osservatorio Astronomico di Roma \\
              via dell'Osservatorio, I-00040 Monteporzio, Italy}

   \date{received ............ ; accepted .............. }

   \maketitle
\markboth {P. Mazzotta et al.: Ionization balance for optically 
thin plasmas} 
{}
\newpage
   \begin{abstract}

We present in this paper new and updated calculations of the 
ionization equilibrium for all the elements from H to Ni.
We   collected for these elements all
 the data available in the literature for the ionization and
radiative plus  dielectronic 
recombination rates. In  particular, the dielectronic rates have 
been fitted with a single  formula and the related coefficients are 
tabulated. 
Our results are compared  with previous works.

   \end{abstract}

\section{Introduction}
Hot plasmas are present in the universe in a variety of astrophysical 
systems,  from 
stellar coronae to the intergalactic medium in clusters of galaxies.
  The good quality data obtained in the last few decades and up to date
from space observatories like EINSTEIN, ROSAT, ASCA and SAX
 allowed to study in many details the physical condition of several 
astrophysical plasmas. 
This activity will have full bloom with the next coming X-ray space
missions, like AXAF,  XMM and ASTRO-E. \\
To determine the relevant physical parameters describing both 
astrophysical plasmas and laboratory  plasmas, 
i.e. electron temperature, density distribution, ion and element 
abundances, we need to compare observed data with a theoretical 
spectral model.\\
Low-density high-temperature astrophysical and laboratory 
plasmas  are generally 
not in a local thermodynamic equilibrium.
To determine the ionization state we 
need to consider in great details the individual collisional 
and radiative ionization and recombination processes.
For a detailed 
discussion of the ionization and recombination processes, see e.g. 
De Michelis \& Mattioli (1981), Mewe (1988, 1990) and Raymond (1988). \\
In the last two decades  various optically thin 
plasma models have been worked out 
[see {
e.g.} \RS (1977), \mew (1972-1986), \rlmf (1990-1991) (hereafter LM) 
, Sutherland 
and Dopita 
(1993)].
 Generally speaking, the main difference among the various
 codes in which the most abundant astrophysical elements 
are considered, lies 
in  the line emission 
calculations , i.e. both in the number of lines and in the atomic data  
considered.
From the early works up to date there have been  extensive  
improvements in the calculation of atomic parameters and, at the same 
time, new X-ray observatories 
 are  able to get the spectrum of a X-ray source 
with better and better energy resolution.
It comes out the great importance of producing reliable  X-ray plasma codes 
by using the most recent atomic data  as well as  by improving 
the algorithms for continuum and line emission processes.\\
The aim of this work is to compute  the ionization balance for the 435  
atoms and 
ions, from H (Z=1) to Ni (Z=28), for plasma temperatures in the range  
$T=0.001 \div 100$
 keV using the most recent data for the ionization and 
recombination rates.\\
Here, we collected all these data making  a critical review of the 
existing works
and a detailed comparison among different sources for the available data. 
In a forthcoming paper (Mazzotta et al. 1998) 
we will use the ionization equilibrium to compute the 
continuum and the line radiation emissions.\\

The plan of the paper is the following:
in section 2 we describe briefly the data used for the 
ionization and 
radiative recombination rates while
in section 3 we discuss,  in details,  the data adopted for the dielectronic 
recombination rates that we fitted with a single  analytic formula. The 
rate
coefficients together with the ionization balance calculations are 
given in tables. In the last section  we compare our results for the 
ionization equilibrium with those obtained in previous works.


\section { Collisional ionization and radiative recombination  rate}
\par
\subsection {Collisional and auto-ionization rate}
To describe the ionization processes we refer to the work of Arnaud and
Rothenflug (1985) and to the updating for the Fe ions of Arnaud and Raymond 
(1992) (hereafter AR85 and AR respectively).
The contributions to the ionization rates\footnote{
The data are available
in electronic form at the following web address: \\
http://www.pa.uky.edu/$^{\sim}$verner/fortran.html}
 are given for all the ions of
H, He, C, N, O, Ne, Na, Mg, Al, Si, S, Ar, Ca and Ni 
from different atomic subshells separately.
The direct ionization rate coefficients
versus temperature $T$ are given by: 
\begin{equation}
    C_{DI}={6.69 \times 10^{7}\over (kT)^{3/2}} 
{\sum_{j} {exp(-x_{j})\over x_{j}}} F(x_{j}) \ \ \ [\mbox{cm}^3/\mbox{s}]
\label{eq2a}
\end{equation}

\begin{equation}
F(x_{j})= A_{j}[1-x_{j}f_{1}(x_{j})]+
\label{eq2b}
\end{equation}
\[
+b{j}[1+x_{j}-x_{j}(2+x_{j})f_1(x_{j})]
C_jf_1(x_{j})+D_jx_{j}f_2(x_{j}). \]
\noindent
The summation is performed over the subshells j of the ionizing ion with
\begin{equation}
x_{j} ={I_j\over kT}, \ \ \ f_1(x)=e^x \int_{1}^{\infty} {dt \over t}
e^{-tx},
\label{eq2c}
\end {equation}
\[
f_2(x) =e^x \int_{1}^{\infty} {dt \over t}
e^{-tx}\ln (t), \] 
\noindent 
and where $kT$ and $I_{j}$ are in eV.
The values parameters $A_{j}$, $B_{j}$, 
$C_{j}$ and $D_{j}$ are given in AR85 and AR.
Following AR85 and AR we also take into 
to account the excitation-autoionization (hereafter EA ) contribution
of ions in the ground state. This is a good approximation for low-density 
plasmas
(as in supernova remnants or clusters of galaxies) because the lifetime of 
excited states is small as compared to the mean collision time. 
The other ions not included in the AR85 work are calculated
 by interpolation
 or extrapolation along the isosequence.

\vskip 0.5truecm
\subsection { Radiative recombination rate }

For the radiative recombination rates$^1$ 
 the calculations of
Shull and Van Steenberg (1982) (hereafter SV) were used for some of the most 
abundant astrophysical elements (Mg, Si, S, Ar, Ca, Fe,
 Ni).   
They give the fitting parameters  $A$ and $\eta$ 
for the following formula:

\begin{equation}
	 \alpha_{r}=A{\left( {T\over 10^4 K} \right)^{\eta}}
	 \ \ \ [\mbox{cm}^3/\mbox{s}],
     	\label{eq16}
\end{equation}
\noindent with the electron temperature  $T$ in eV.
LM extrapolated these calculations also to other astrophysical less 
abundant  elements.
For Fe~XV--Fe~XXIV we used  the formula  of
 AR:

\begin{equation}
	 \alpha_{r}=A{\left( {T\over 10^4 K} \right)^{\alpha-\beta 
\log_{10}\left( {T\over 10^4 K} \right) }}
        \ \ \ [\mbox{cm}^3/\mbox{s}],
	\label{eq6}
\end{equation}
\noindent 
where the electron temperature $T$ is in Kelvin and 
the fitting parameters $A, \alpha, 
\beta $ are given by AR
in tabular form.
For the  H-like, He-like, Li-like and Na-like isosequences we used the 
new calculations, in the framework of the 
opacity project, of Verner \& Ferland (1996). They fit the data by
the following  formula:
\begin{equation}
	 \alpha_{r}=A\left[{\sqrt{T\over T_0}\left(1+ \sqrt
{T\over T_0} \right)^{1-b}
\left(1+ \sqrt{T\over T_1} \right)^{1+b}}\right]^{-1}	\label{eq7}
\end{equation}
\[ 
\ \ \ \ \ \ \ \ \ \ \ \ \ \ \ \ \ \ \ \ \ \ \ \ \ \ \ \ \ \ 
\ \ \ \ \ \ \ \ \ \ \ \ \ \ \ \ \ \ \ \ \ \ \ \ \ \ \ \ \ \ 
\ \ [\mbox{cm}^3/\mbox{s}],
 \]
with  the electron temperature $T$  in Kelvin, and  
  the fitting parameters  $A,b, T_{0}, 
T_{1} $ are given 
for all the ions from H through Zn. This last formula is   valid in a 
wide range of temperatures, from T=$2.5 \cdot 10^{-4}$ eV  to T=$100$ keV. 
We also use the same formula 
for C, N, O and Ne ions   with the fitting parameters
 of P\'equignot et al. (1991).\\


\section { Dielectronic recombination rate}
 Burgess (1965) pointed out  the relevance of the  
dielectronic recombination (hereafter DR) process  especially in the temperature region of 
maximum ion abundance.   
The collected dielectronic recombination rates are calculated in the 
zero electron density limit, which is a valid approximation 
for most   of the astrophysical 
plasmas.\\ 
In the following, as in  AR,  we 
discuss the dielectronic recombination 
rates by isosequence and we explicitly declare  when we 
 have chosen different data from them. \\
 All the available data were fitted, when 
needed, with the following formula:  
\begin{equation}
	\alpha_{\mbox d}=
        {1\over T^{3/2}}\sum_{j=1}^{4}c_{j}\exp\left(-{E_{i} \over 
	T}\right) \ \ \ [{\mbox{cm}}^{3}/\mbox{s}],
	\label{eq8}
\end{equation}
\noindent
where $T$  and $E_{i}$ are given in eV and $c_{i}$ in ${\mbox{cm}}^{3}
{\mbox {s}}^{-1}$.
The coefficients $c_{i}$ and $E_{i}$  are given in ~Table 1.\\
{\it H-like - }
From C to Ni we used the data of \ra (hereafter M) and \rb  which for the same 
elements are in good agreement.
For the elements not explicitly calculated  we interpolated the DR rates along 
the isosequence. 
For Be IV and B V we used the data of Pindzola \& Badnell (1992) that are
a factor 2 less then those from  previous calculations of Shore (1969) 
with the corrected values  by  Burgess and Tworkowski (1976).
Moreover we divided by a factor 2 also the DR coefficients for HeII and Li~III
by Shore (1969).\\
{\it {He-like -}}
We fitted with our formula the calculations of
Chen (1986a)
 that are in very good agreement with Nielsen (1986) and Karim \& Bhalla
(1989).
For  C V we adopt Chen's (1988b) calculation in which the Coster-Kroning 
transitions are taken into account that  are negligible for higher Z   
elements. For BeIII and B IV we adopted Pindzola \& Badnell (1992).\\
{\it {Li-like -}}
We adopt the DR calculations of \re. The total DR rate coefficient for the
11 lithium-like ions were calculated using the distorted wave technique and
the multi-configuration Dirac-Fock method. The discrepancies with 
non-relativistic calculations of  Roszman (1987)
are more relevant for highly 
charged medium- and high-Z ions. 
For the other 
ions we have interpolated 
along the isosequence. For BeII and B III we have adopted 
the DR calculations of Pindzola \& Badnell (1992).   \\
 {\it {Be-like -}}
We adopt  Badnell's (1987) calculations 
which for high Z agree well with Chen \& Crasemann (1988). For O V and B II 
we use, respectively, the data of Badnell \& Pindzola (1989) and 
Pindzola \& Badnell (1992). \\
  {\it {B-like -}}
Recent calculations of Nahar (1995) and errata from  Nahar (1996a),
 in the framework of a unified treatment of 
electron-ion recombination, show good agreement with low-T DR rates by 
Nussbaumer and Storey (1983), and high-T DR rates by Jacobs et al. (1979)
 for  the ions from C to Al, and for Si and  S.
As for Fe, we use the data of AR. 
For most of the other ions calculations from SV, LM and M are available.
For the other ions   
we used the Burgess (1965) general formula, corrected by Merts et al. 
(1976) (hereafter Burgess-Merts formula, BM),  using the 
most recent data for oscillator strengths and energy transitions.\\ 
  Iron is the most 
investigated element so that in the following we always adopt the 
criteria, except when more recent calculations are available, to 
normalize the DR calculation rates to the
 Fe calculations of AR. 
These criteria can be  also justified by the fact that, as  pointed out in  
LM (1991) and Hahn (1991), most of the coefficients show a regular trend 
along the same isosequence.\\ 
 {\it {C-like -}} {\it {N-like -}}
From Ar to Ni we  followed the same procedure  described for 
B-like ions. For the other elements, except for O III for which we used the 
data from Badnell and Pindzola (1989), we used the data of Jacobs et al. 
(1977a, 1977b, 1979, 1980)
and LM (1991).\\ 
  {\it {O-like -}}  {\it {F-like -}}
From Ar to Ni we adopt the recent calculations  of 
Dasgupta \& Whitney (1990,1994)
 that for F-like 
ions are in good agreement with Chen (1988a)
 calculations. For the other elements 
we used the calculations of SV.\\
 {\it {Ne-like -}} 
We use the calculations of \rd,  \ro and for Na II the 
calculation of SV. For P VI and Cl VIII we adopted the data of LM(1991)
multiplied by a factor 4 to take into to account the results of 
Romanik (1988) and  Chen (1986b) for adjacient ions.\\
 {\it {Na-like -}}  {\it {Mg-like -}}
As it has been indicated by Zhdanov (1982), Dube \& LaGatutta (1987), Dube,
Rosoanaivo \& Hahn (1985), 
at high temperature the inner shell transitions of 
the form 2s-3d become important, and this effect is more important for 
medium and high-Z elements. We follow the procedure of Mattioli (1988) 
 that for Fe
is in very good agreement with AR. For the other elements we adopt the 
calculations of Jacobs and LM taking into account the results obtained for 
Mg II and Si IV from LaGattuta \& Hahn (1982,1984). For S VI we adopt 
the data of Badnell (1991)\\  
 {\it {Al-like -}} to {\it {Co-like -}}
  For all these isosequences, except for  Fe, detailed or recent 
calculations are
not available so that we used data from SV, LM,
M as well as the general BM formula. Up to the Mn-like isosequence 
we renormalize the results to the work of AR. \\
\noindent	
We mention that recently new calculations have been made on the 
recombination of Fe II (Nahar et al. 1997) and Fe IV (Nahar 1996b).
In the case of Fe II the DR rate is a factor 2 above the AR curve
 whereas for Fe IV the Nahar results are an order of magnitude below the AR 
curve although Nahar claims a 10-30\% accuracy.
We do not know the cause of the discrepancies but for consistency we prefer
to use the AR results. 
In the last years Teng et al. (1994a, 1994b, 1994c, 1994 d, 1996)
also applied fitting formula for several isoelectronic sequence i.e.
 H, He, Li, F and Ne. In general our results are within a 10\% in agreement 
with those calculations essentially because we have fitted and/or 
interpolated the same data, but in some case their fitting formula yield 
divergences in  the dielectronic rates.

\section { Results for the coronal equilibrium}

\begin{figure}
\includegraphics[width=8.8cm]{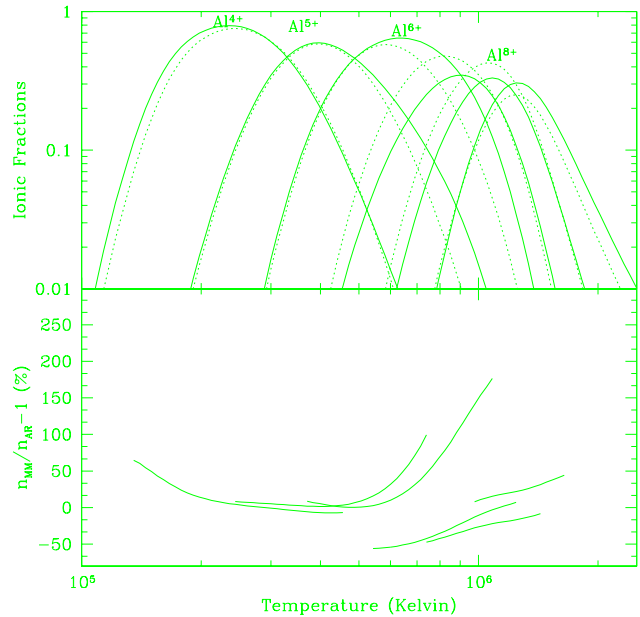}
\caption[]{-- {\it Upper panel}: 
Ionic fraction vs. temperature for Al ions, from 
Al V to Al X. 
    Solid curves: present work; dashed curves:  AR85.} 
{\it Lower panel}: percent variations in the ionic abundance
fractions in the 
  present work with respect to that of AR85.
For each ion, the percent variations are evaluated only for a range of
temperatures in which the respective ionic fractions are  $> 10^{-1}$.
\end{figure}

\begin{figure}
\includegraphics[width=8.8cm]{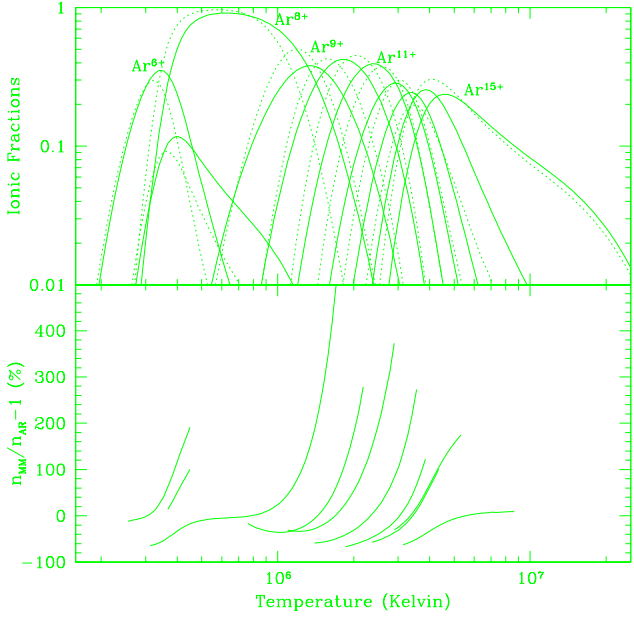}
\caption[]{-- Same as Fig.1 but for Ar ions, from Ar~VII to Ar~XXVI.}
\end{figure}

\begin{figure}
\includegraphics[width=8.8cm]{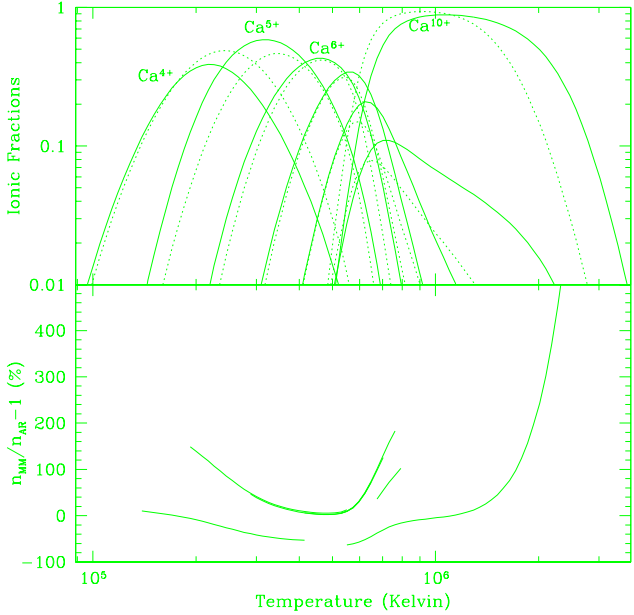}
\caption[]{-- Same as Fig.1 but for Ca ions, from Ca~V to Ca~XI.}
\end{figure}

\begin{figure}
\includegraphics[width=8.8cm]{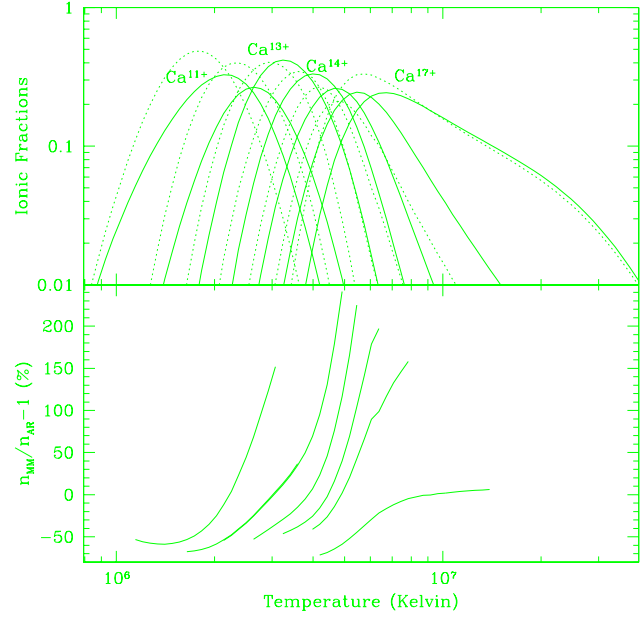}
\caption[]{-- Same as Fig.1 but for Ca ions, from Ca~XII to Ca~XVIII.}
\end{figure}

\begin{figure}
\includegraphics[width=8.8cm]{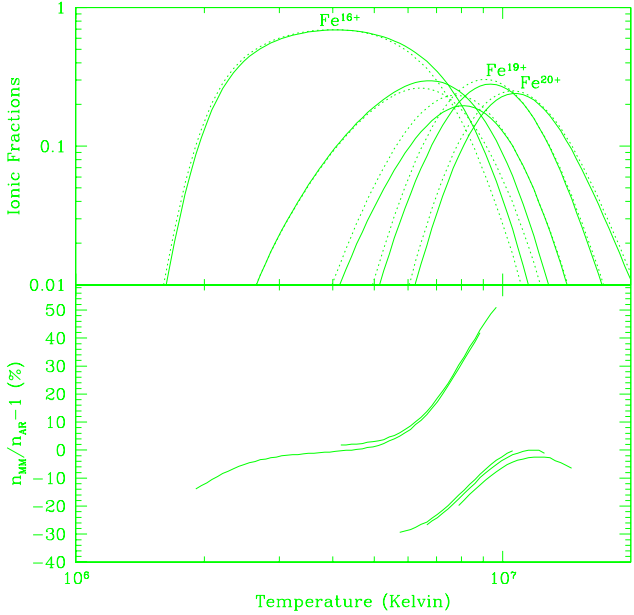}
\caption[]{-- {\it Upper panel}: 
Ionic fraction vs. temperature  for Fe  ions, from Fe~XVII to Fe~XXI.
    Solid curves: present work; dashed curves:  AR.} 
{\it Lower panel}: percent variations in the ionic abundance
fractions in
  present work with respect to  AR.
For each ion the percent variations are evaluated only for a range of
temperatures in which the respective ionic fractions are  $> 10^{-1}$.
\end{figure}

\begin{figure}
\includegraphics[width=8.8cm]{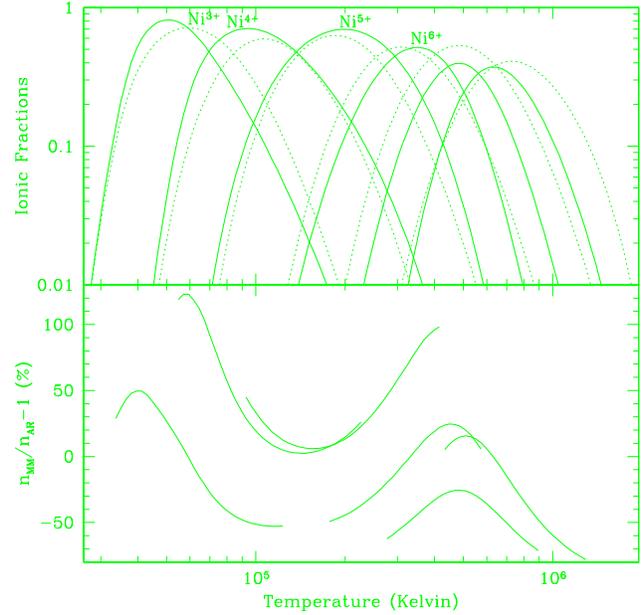}
\caption[]{-- Same as Fig.1 but for Ni  ions, from Ni~IV to Ni~IX.}
\end{figure}

\begin{figure}
\includegraphics[width=8.8cm]{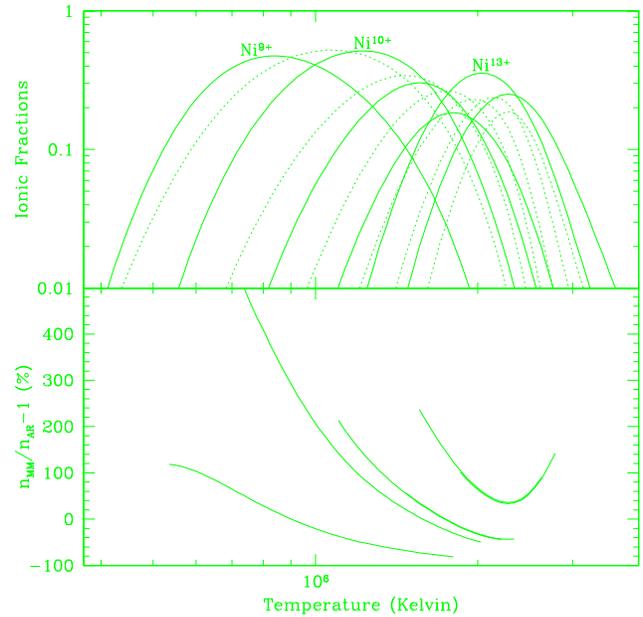}
\caption[]{-- Same as Fig.1 but for Ni  ions, from Ni~X to Ni~XV.}
\end{figure}

\begin{figure}
\includegraphics[width=8.8cm]{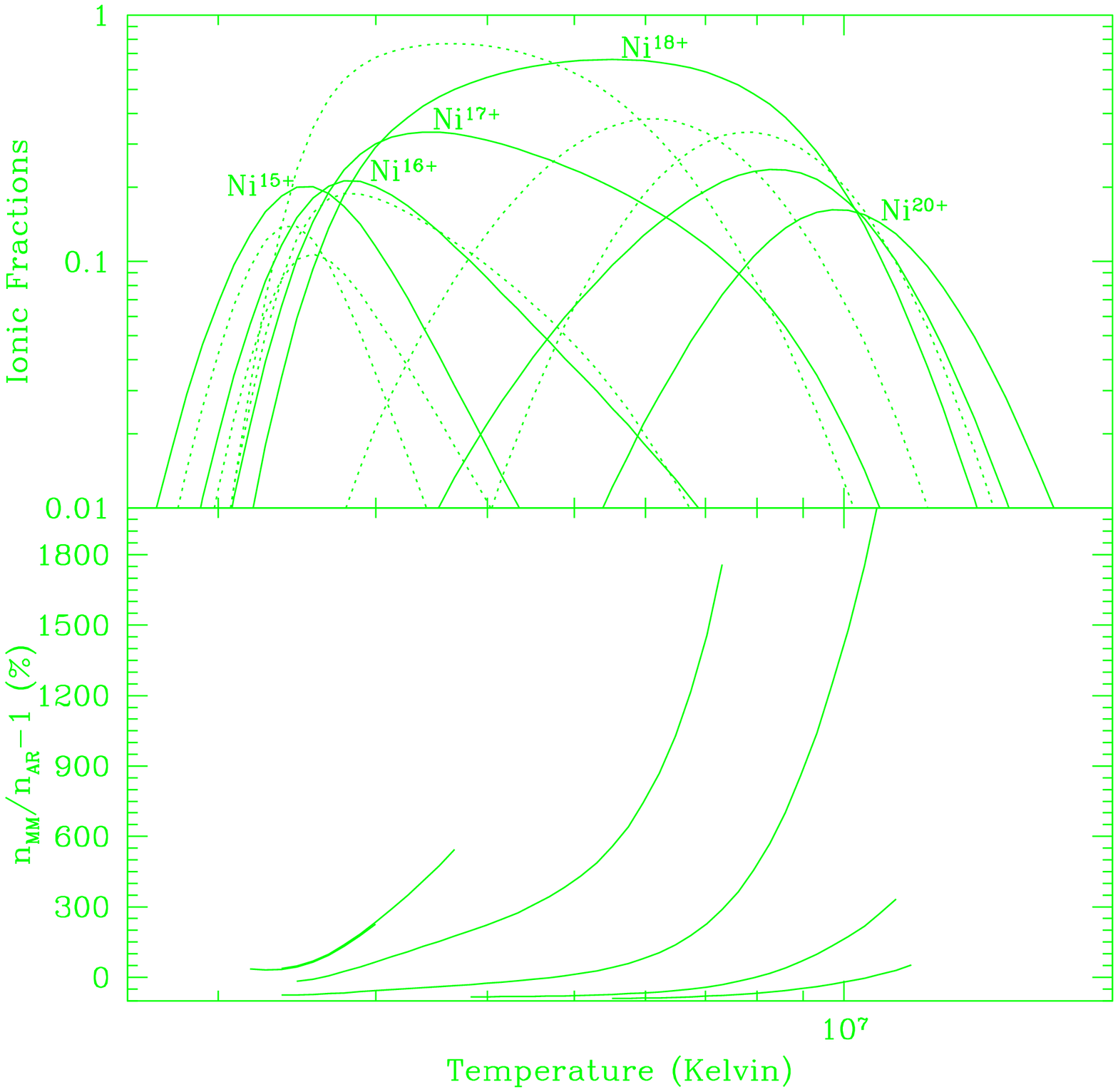}
\caption[]{-- Same as Fig.1 but for Ni  ions, from Ni~XVI to Ni~XXI.}
\end{figure}

\begin{figure}
\includegraphics[width=8.8cm]{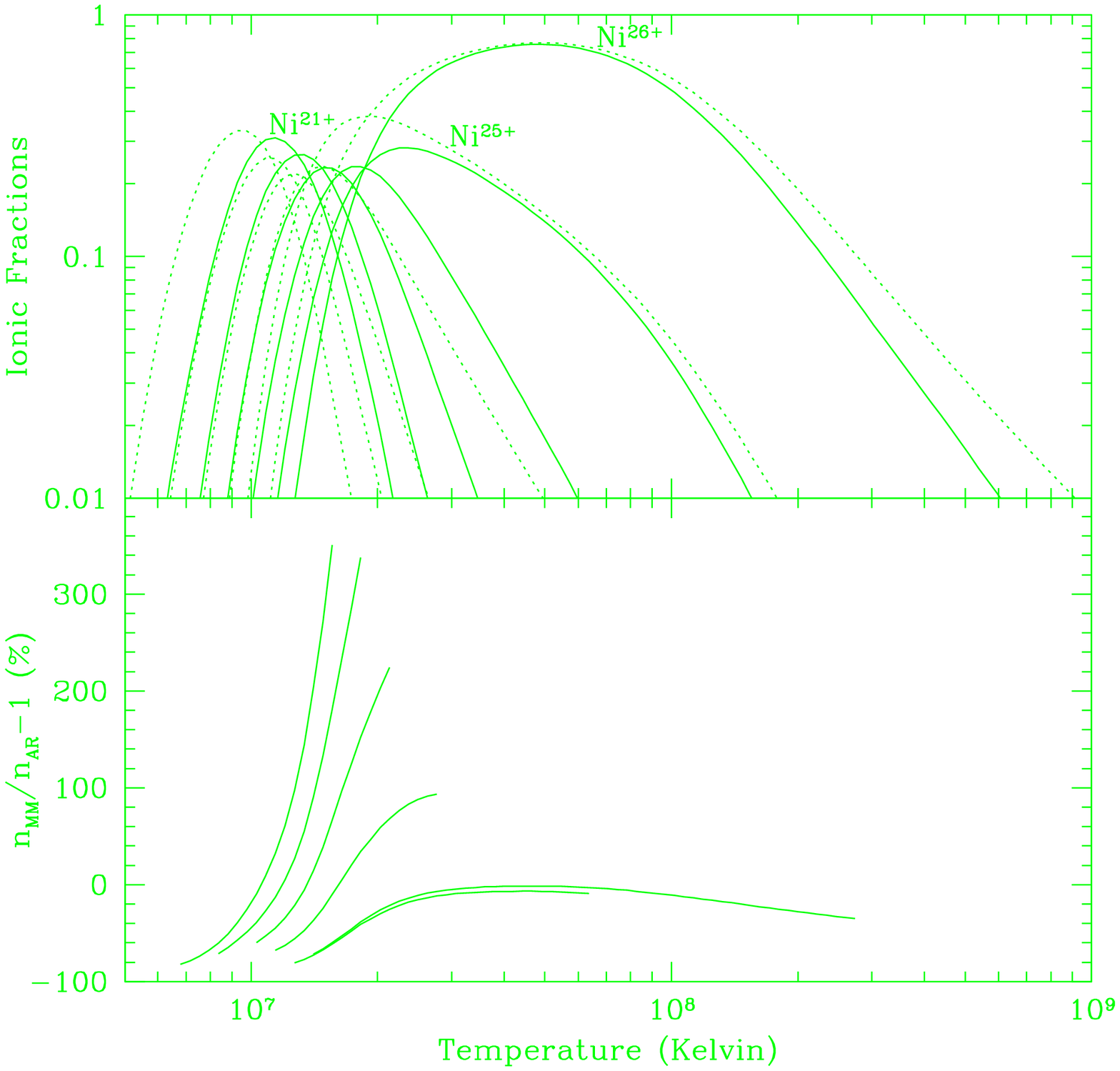}
\caption[]{-- Same as Fig.1 but for Ni  ions, from Ni~XXII to Ni~XXVII.}
\end{figure}

\begin{figure}
\includegraphics[width=8.8cm]{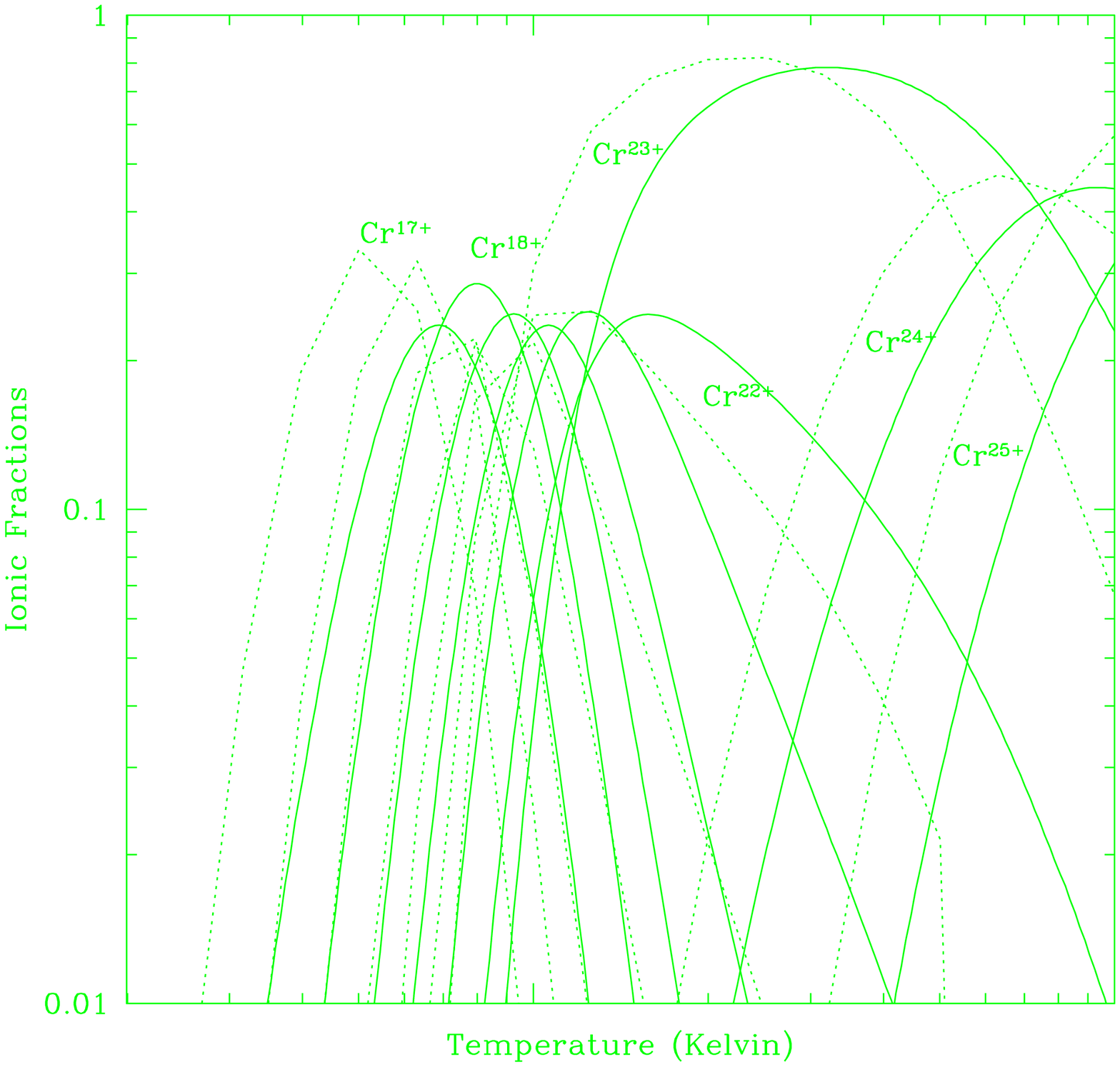}
\caption[]{-- Ionic fraction vs. temperature Cr ions, from 
Cr~XVIII to Cr~XXVI.  
Solid curves: present work; dashed curves:  \rlmf(1991).}
\end{figure}

In Table 2 we give the ionization fraction $X_{i}$, for the 28 
elements.
In these calculations  we  included only the collisional processes
in the low-density case.
We compared our ionic abundances fractions with those given by 
AR85 (AR for the iron).
We found a very good agreement or differences less then 10\% for all 
the atoms and ions of 
H, He, N, Ne, Na, Mg, Si.
For  C~II, C~III, O~III, O~IV, O~V we found differences 
up to 50\% near the peaks of maximum ionic abundance,
but good agreement for the other ions of these elements.
In figure 1 we compare our ionic fraction  
 for the less  ionized ions of Al with those  of AR85.
In the same figure we report, for each ion,
 also the percent variation  
near the peak of maximum ionic abundance. We can note that our ionic 
abundance fraction at $T\approx 2\times 10^{5} $K is a factor of 2
greater than those of AR85.
In figures 2 to 4 we report the comparisons for some Ar
and Ca ions.
In those cases we found, depending on the temperature and on
the ions considered, differences up to 400\%. 
The curves obtained for the iron ions 
are both or the same or in very good 
agreement with those of AR except for the ions
from Fe~XVII to Fe~XXI due to the updating of Fe XIX dielectronic recombination
 rate, as shown by figure 5, where we still have small
variations.
In figures 6 to 9 we compare the curves for some Ni ions and
 we found  substantially differences with respect to AR85. 
If we fix the temperature at  which the ion abundance 
curves reach their maxima,
we can see that generally the big differences found
   are due to  a
shift, depending on the  considered ion, 
to higher or lower temperature with respect to the other considered  case.\\
We compered, also, our ionic abundance curves for the less abundant 
astrophysical elements with those of \rlmf (1991) and found again 
significant differences, depending on the temperature and ion considered.
In figure 10, as an example, we show the comparison of our ionic fraction 
curves for the most ionized ions 
of Cr with those  of \rlmf (1991) and we can observe that our curves are shifted to higher temperatures.


\section { Discussion}

The main 
goal of this work was also to  collect all available
 data on the collisional
ionization, radiative and dielectronic recombination rates data.
Most of  this data come out from new advanced computational calculations.   
About the ionization rates we would like to remark that recently Kato et al.
(1991) 
reviewed and compared the rates obtained with several empirical formulae
used by Lotz (1967, 1968), by the  Belfast group
 [Bell et al. (1983) and Lennon et al. (1988)], by
  AR85 and AR, and by Pindzola et al. (1987).
They conclude that generally 
the rate coefficient derived on the basis of experimental 
data of cross sections are in good agreement among the various authors.
Otherwise when no experimental data are available only AR85 and AR included 
the EA process. This reflects in a difference at  high temperature  
for some isosequences
(e.g., Na-like, Li-like, and Mg- to Ar-like sequences).
Kato et al. (1991) 
also remark that  for some ions the Belfast group rates
are  lower by a factor 1.626 due to a misprint on the Younger (1982)
table.  
Based on the Kato results, Voronov (1997) refitted the data of the Belfast
group with a very simple formula for the total ionization rates.
Instead of its simplicity we do not use the Voronov data for the following
reason:\
i) Voronov give a fit only for the total ionization rate. This formula is
valid as long as we consider the case of ionization equilibrium plasmas.
Under non-equilibrium conditions (such as apply to transient
plasmas which are
present e.g., in supernova remnants and solar/stellar flares) 
 inner-shell ionization
may play an important role, both in the determination of the
ionization balance and in the formation of fluorescent lines.
Then it can be important in certain cases (e.g., for the Be- and B-like 
sequences) to know the contribution from different atomic subshells separately
(Mewe private communication).\
 ii) we checked the Voronov results and we saw that, instead of his claim,
 some curves are in agreement with the original
data of the Belfast group without the correction factor of 1.626
(which implies that in these cases his results are just a
factor of 1.626 below the corresponding AR results, see e.g. Kato et
al. (1991) for Fe XXII to Fe XXVI).
So, for some ions, the Voronov ionization rates 
are uncorrected.\
We would like to remark that the ionization rates for Ni I to Ni X are probably 
wrong
by large factors due to a large underestimate of the Excitation autoionization 
process in the AR85 (Arnaud private communication). In a forthcoming paper (Kaastra et al. 1998) we 
intend to update the ionization rate including EA, resonance excitation double 
autoionization and direct multiple ionization.

We believe that at present 
our atomic data collection represents
 the state of the art in such calculations.
An important point of this work was  to  
reorganize all the available data  with a few and  simple fitting
 formulae in order
to use them in a numerical modeling of plasma processes.\\  
Obviously, the differences in the ionization fraction
between our work and the previous ones, 
reflects directly in the computed line power emissions.
In particular, our  preliminary calculations show
sensitive  differences, depending on
the selected plasma temperature,  in the computation
of the total line radiation emission with respect to  previous similar work.
 We will fully address  these issues in a forthcoming  paper.

\par

{\it Note:} Our ionization and recombination rate computation codes are 
available on request.

\begin{acknowledgements}
We are grateful to R. Mewe, for  helpful comments
 and suggestions. We are indebted to M. Mattioli who has pointed out to us
some mistakes in the data.

\end{acknowledgements}

\newpage
\onecolumn
\input {tab_1.tex}
\newpage
\input {tab_2.tex}

\end{document}

%% file: tab_1.tex
 \pagestyle{empty}
 \centerline{Table 1. Fitting coefficients for dielectronic recombination rates}
 {\scriptsize
\begin{tabular}{||l|r r r r|r r r r||}
     \hline
     \hline
       Ions & $c_{1}$ & $c_{2}$ & $c_{3}$ &   $c_{4}$ & $E_{1}$ & $E_{2}$ &$E_{3}
 $ &  $E_{4}$ \\
       \hline
 HeII    
 &0.22400E-08&........... &...........&........... &   39.70&........  &........  &........  \\
 LiII    
 &0.13200E-08&........... &...........&........... &   64.70&........  &........  &........  \\
 LiIII   
 &0.95990E-08&........... &...........&........... &   91.80&........  &........  &........  \\
 BeII    
 &0.52662E-09&........... &...........&........... &    4.01&........  &........  &........  \\
 BeIII   
 &0.12621E-08&........... &...........&........... &  118.51&........  &........  &........  \\
 BeIV    
 &0.87198E-08&........... &...........&........... &  159.13&........  &........  &........  \\
 B II    
 &0.13602E-08&........... &...........&........... &    9.95&........  &........  &........  \\
\hline
 \hline
 \end{tabular}}
 \newpage
 \centerline{Table 1. -- (continued)}
 {\scriptsize
 \begin{tabular}{||l|r r r r|r r r r||}
     \hline
     \hline
       Ions & $c_{1}$ & $c_{2}$ & $c_{3}$ &   $c_{4}$ & $E_{1}$ & $E_{2}$ &$E_{3}
 $ &  $E_{4}$ \\
       \hline
AlVIII  
 &0.15221E-06&0.56355E-07&...........  &...........  &  168.98&   36.54&........  &........    \\
 AlIX    
 &0.72670E-10&0.55499E-09&0.19795E-07&0.67984E-07&    0.52&    3.61&   28.71&  122.65  \\
 AlX     
 &0.23354E-07&0.17971E-07&0.10436E-07&...........  &   57.44&  379.30&   16.58&........    \\
 AlXI    
 &0.47398E-08&0.55815E-07&0.91269E-08&...........  &   12.20&  419.18&   31.14&........    \\
 AlXII   
 &0.17562E-06&........... &...........&........... & 1386.10&........  &........  &........  \\
 AlXIII  
 &0.10226E-06&........... &...........&........... & 1489.60&........  &........  &........  \\
 SiII    
 &0.48978E-11&0.60799E-09&...........  &...........  &    2.05&    9.68&........  &........    \\
\hline
 \hline
 \end{tabular}}
 \newpage
 \centerline{Table 1. -- (continued)}
 {\scriptsize
 \begin{tabular}{||l|r r r r|r r r r||}
     \hline
     \hline
       Ions & $c_{1}$ & $c_{2}$ & $c_{3}$ &   $c_{4}$ & $E_{1}$ & $E_{2}$ &$E_{3}
 $ &  $E_{4}$ \\
       \hline
K II    
 &0.18972E-07&........... &...........&........... &   44.41&........  &........  &........  \\
 K III   
 &0.87035E-08&........... &...........&........... &   32.74&........  &........  &........  \\
 K IV    
 &0.37559E-07&........... &...........&........... &   30.42&........  &........  &........  \\
 K V     
 &0.24985E-07&........... &...........&........... &   26.73&........  &........  &........  \\
 K VI    
 &0.87476E-07&........... &...........&........... &   24.81&........  &........  &........  \\
 K VII   
 &0.92957E-07&........... &...........&........... &   24.73&........  &........  &........  \\
 K VIII  
 &0.69297E-07&........... &...........&........... &   20.23&........  &........  &........  \\
\hline
 \hline
 \end{tabular}}
 \newpage
 \centerline{Table 1. -- (continued)}
 {\scriptsize
 \begin{tabular}{||l|r r r r|r r r r||}
     \hline
     \hline
       Ions & $c_{1}$ & $c_{2}$ & $c_{3}$ &   $c_{4}$ & $E_{1}$ & $E_{2}$ &$E_{3}
 $ &  $E_{4}$ \\
       \hline
V V     
 &0.61416E-07&........... &...........&........... &   50.94&........  &........  &........  \\
 V VI    
 &0.73091E-07&........... &...........&........... &   46.56&........  &........  &........  \\
 V VII   
 &0.18490E-06&........... &...........&........... &   53.54&........  &........  &........  \\
 V VIII  
 &0.18439E-06&........... &...........&........... &   51.34&........  &........  &........  \\
 V IX    
 &0.99104E-07&........... &...........&........... &   45.86&........  &........  &........  \\
 V X     
 &0.12085E-06&0.66932E-07&...........  &...........  &   37.03&   72.23&........  &........    \\
 V XI    
 &0.16869E-06&........... &...........&........... &   32.99&........  &........  &........  \\
\hline
 \hline
 \end{tabular}}
 \newpage
 \centerline{Table 1. -- (continued)}
 {\scriptsize
 \begin{tabular}{||l|r r r r|r r r r||}
     \hline
     \hline
       Ions & $c_{1}$ & $c_{2}$ & $c_{3}$ &   $c_{4}$ & $E_{1}$ & $E_{2}$ &$E_{3}
 $ &  $E_{4}$ \\
       \hline
FeXVII  
 &0.45040E-06&0.63367E-06&...........  &...........  &  479.98&  802.63&........  &........    \\
 FeXVIII 
 &0.20250E-08&0.26894E-07&0.14487E-06&0.15368E-05&   22.50&  117.00&  341.00&  683.00  \\
 FeXIX   
 &0.17369E-06&0.98644E-08&0.17365E-06&0.14842E-05&  411.02&   50.61&  410.90&  780.02  \\
 FeXX    
 &0.12807E-07&0.57389E-07&0.72517E-07&0.59150E-06&   23.70&   85.10&  329.00&  787.00  \\
 FeXXI   
 &0.14808E-07&0.76279E-07&0.63232E-07&0.98451E-06&   13.20&   66.60&  297.00&  714.00  \\
 FeXXII  
 &0.74358E-09&0.10325E-06&0.15368E-06&0.72997E-06&   39.10&   80.30&  392.00&  919.00  \\
 FeXXIII 
 &0.40660E-07&0.83362E-07&0.49659E-06&...........  &   29.03&  113.33&  765.77&........    \\
\hline
 \hline
 \end{tabular}}

%% file: tab_2.tex
 \centerline{Table 2. - Ionization equilibria }
  
 {\tiny
 \begin{tabular}{ c c c c c}                                                    
 \hline \hline
 &$\log_{10}$(T)&
 H I     &
 H II    & \\
 \hline
&4.00&
0.001&
2.659& \\
&4.10&
0.031&
1.161& \\
&4.20&
0.359&
0.250& \\
&4.30&
1.151&
0.032& \\
&4.40&
1.954&
0.005& \\
&4.50&
2.636&
0.001& \\
&4.60&
3.208&
0.000& \\
&4.70&
3.691&
0.000& \\
&4.80&
4.103&
0.000& \\
&4.90&
4.458&
0.000& \\
&5.00&
4.768&
0.000& \\
&5.10&
 ...&
0.000& \\
&5.20&
 ...&
0.000& \\
&5.30&
 ...&
0.000& \\
&5.40&
 ...&
0.000& \\
&5.50&
 ...&
0.000& \\
&5.60&
 ...&
0.000& \\
&5.70&
 ...&
0.000& \\
&5.80&
 ...&
0.000& \\
     \hline
 \end{tabular}
 }
 \hspace{0.5cm}
  
 {\tiny
 \begin{tabular}{ c c c c c c}                                                  
 \hline \hline
 &$\log_{10}$(T)&
 HeI     &
 HeII    &
 HeIII   & \\
 \hline
&4.00&
0.000&
 ...&
 ...\\
&4.10&
0.000&
 ...&
 ...\\
&4.20&
0.000&
3.993&
 ...\\
&4.30&
0.002&
2.258&
 ...\\
&4.40&
0.058&
0.904&
 ...\\
&4.50&
0.478&
0.175&
 ...\\
&4.60&
1.263&
0.024&
3.755& \\
&4.70&
1.970&
0.007&
2.202& \\
&4.80&
2.455&
0.048&
0.997& \\
&4.90&
2.934&
0.335&
0.270& \\
&5.00&
3.640&
0.948&
0.052& \\
&5.10&
4.369&
1.592&
0.011& \\
&5.20&
 ...&
2.153&
0.003& \\
&5.30&
 ...&
2.629&
0.001& \\
&5.40&
 ...&
3.032&
0.000& \\
&5.50&
 ...&
3.377&
0.000& \\
&5.60&
 ...&
3.676&
0.000& \\
&5.70&
 ...&
3.937&
0.000& \\
&5.80&
 ...&
4.167&
0.000& \\
     \hline
 \end{tabular}
 }
 \newpage
 \centerline{Table 2. -- (continued) }
  
 {\tiny
 \begin{tabular}{ c c c c c c c}                                                
 \hline \hline
 &$\log_{10}$(T)&
 LiI     &
 LiII    &
 LiIII   &
 LiIV    & \\
 \hline
&4.00&
2.647&
0.001&
 ...&
 ...\\
&4.10&
3.323&
0.000&
 ...&
 ...\\
&4.20&
3.884&
0.000&
 ...&
 ...\\
&4.30&
4.352&
0.000&
 ...&
 ...\\
&4.40&
4.745&
0.000&
 ...&
 ...\\
&4.50&
 ...&
0.000&
 ...&
 ...\\
&4.60&
 ...&
0.000&
 ...&
 ...\\
&4.70&
 ...&
0.000&
4.560&
 ...\\
&4.80&
 ...&
0.001&
2.870&
 ...\\
&4.90&
 ...&
0.014&
1.513&
 ...\\
&5.00&
 ...&
0.148&
0.540&
4.183& \\
&5.10&
 ...&
0.595&
0.130&
2.384& \\
&5.20&
 ...&
1.100&
0.067&
1.194& \\
&5.30&
 ...&
1.571&
0.222&
0.428& \\
&5.40&
 ...&
2.210&
0.661&
0.110& \\
&5.50&
 ...&
2.922&
1.205&
0.029& \\
&5.60&
 ...&
3.587&
1.707&
0.009& \\
&5.70&
 ...&
4.185&
2.141&
0.003& \\
&5.80&
 ...&
4.723&
2.513&
0.001& \\
     \hline
 \end{tabular}
 }
 \hspace{0.5cm}
  
 {\tiny
 \begin{tabular}{ c c c c c c c c}                                              
 \hline \hline
 &$\log_{10}$(T)&
 BeI     &
 BeII    &
 BeIII   &
 BeIV    &
 BeV     & \\
 \hline
&4.00&
0.052&
0.950&
 ...&
 ...&
 ...\\
&4.10&
0.237&
0.376&
3.988&
 ...&
 ...\\
&4.20&
0.620&
0.123&
2.114&
 ...&
 ...\\
&4.30&
1.152&
0.114&
0.793&
 ...&
 ...\\
&4.40&
2.046&
0.541&
0.153&
 ...&
 ...\\
&4.50&
3.195&
1.283&
0.024&
 ...&
 ...\\
&4.60&
4.252&
1.980&
0.005&
 ...&
 ...\\
&4.70&
 ...&
2.568&
0.001&
 ...&
 ...\\
&4.80&
 ...&
3.060&
0.000&
 ...&
 ...\\
&4.90&
 ...&
3.474&
0.000&
 ...&
 ...\\
&5.00&
 ...&
3.824&
0.000&
 ...&
 ...\\
&5.10&
 ...&
4.122&
0.000&
3.499&
 ...\\
&5.20&
 ...&
4.375&
0.003&
2.111&
 ...\\
&5.30&
 ...&
4.596&
0.043&
1.026&
4.404& \\
&5.40&
 ...&
4.895&
0.258&
0.352&
2.480& \\
&5.50&
 ...&
 ...&
0.694&
0.131&
1.241& \\
&5.60&
 ...&
 ...&
1.222&
0.211&
0.489& \\
&5.70&
 ...&
 ...&
1.880&
0.557&
0.149& \\
&5.80&
 ...&
 ...&
2.590&
1.022&
0.045& \\
&5.90&
 ...&
 ...&
3.253&
1.469&
0.015& \\
&6.00&
 ...&
 ...&
3.847&
1.863&
0.006& \\
&6.10&
 ...&
 ...&
4.380&
2.203&
0.003& \\
&6.20&
 ...&
 ...&
4.862&
2.498&
0.001& \\
&6.30&
 ...&
 ...&
 ...&
2.756&
0.001& \\
&6.40&
 ...&
 ...&
 ...&
2.984&
0.000& \\
&6.50&
 ...&
 ...&
 ...&
3.188&
0.000& \\
     \hline
 \end{tabular}
 }
 \newpage
 \centerline{Table 2. -- (continued) }
  
 {\tiny
 \begin{tabular}{ c c c c c c c c c}                                            
 \hline \hline
 &$\log_{10}$(T)&
 B I     &
 B II    &
 B III   &
 B IV    &
 B V     &
 B VI    & \\
 \hline
&4.00&
0.956&
0.051&
 ...&
 ...&
 ...&
 ...\\
&4.10&
1.765&
0.008&
 ...&
 ...&
 ...&
 ...\\
&4.20&
2.225&
0.003&
 ...&
 ...&
 ...&
 ...\\
&4.30&
2.431&
0.002&
3.770&
 ...&
 ...&
 ...\\
&4.40&
2.564&
0.002&
2.565&
 ...&
 ...&
 ...\\
&4.50&
2.697&
0.012&
1.582&
4.531&
 ...&
 ...\\
&4.60&
2.888&
0.074&
0.818&
2.403&
 ...&
 ...\\
&4.70&
3.291&
0.340&
0.391&
0.868&
 ...&
 ...\\
&4.80&
4.231&
1.135&
0.598&
0.171&
 ...&
 ...\\
&4.90&
 ...&
2.241&
1.200&
0.031&
 ...&
 ...\\
&5.00&
 ...&
3.267&
1.789&
0.007&
 ...&
 ...\\
&5.10&
 ...&
4.154&
2.293&
0.002&
 ...&
 ...\\
&5.20&
 ...&
 ...&
2.719&
0.001&
 ...&
 ...\\
&5.30&
 ...&
 ...&
3.078&
0.000&
4.283&
 ...\\
&5.40&
 ...&
 ...&
3.379&
0.001&
2.814&
 ...\\
&5.50&
 ...&
 ...&
3.613&
0.010&
1.637&
 ...\\
&5.60&
 ...&
 ...&
3.802&
0.082&
0.764&
3.183& \\
&5.70&
 ...&
 ...&
4.058&
0.327&
0.293&
1.708& \\
&5.80&
 ...&
 ...&
4.424&
0.732&
0.188&
0.781& \\
&5.90&
 ...&
 ...&
 ...&
1.276&
0.368&
0.285& \\
&6.00&
 ...&
 ...&
 ...&
1.930&
0.739&
0.094& \\
&6.10&
 ...&
 ...&
 ...&
2.581&
1.149&
0.033& \\
&6.20&
 ...&
 ...&
 ...&
3.177&
1.527&
0.013& \\
&6.30&
 ...&
 ...&
 ...&
3.713&
1.858&
0.006& \\
&6.40&
 ...&
 ...&
 ...&
4.197&
2.148&
0.003& \\
&6.50&
 ...&
 ...&
 ...&
4.638&
2.402&
0.002& \\
&6.60&
 ...&
 ...&
 ...&
 ...&
2.626&
0.001& \\
&6.70&
 ...&
 ...&
 ...&
 ...&
2.826&
0.001& \\
&6.80&
 ...&
 ...&
 ...&
 ...&
3.007&
0.000& \\
     \hline
 \end{tabular}
 }
 \hspace{0.5cm}
  
 {\tiny
 \begin{tabular}{ c c c c c c c c c c}                                          
 \hline \hline
 &$\log_{10}$(T)&
 C I     &
 C II    &
 C III   &
 C IV    &
 C V     &
 C VI    &
 C VII   & \\
 \hline
&4.00&
0.018&
1.396&
 ...&
 ...&
 ...&
 ...&
 ...\\
&4.10&
0.212&
0.414&
 ...&
 ...&
 ...&
 ...&
 ...\\
&4.20&
0.736&
0.088&
4.494&
 ...&
 ...&
 ...&
 ...\\
&4.30&
1.252&
0.026&
2.824&
 ...&
 ...&
 ...&
 ...\\
&4.40&
1.639&
0.020&
1.664&
 ...&
 ...&
 ...&
 ...\\
&4.50&
1.974&
0.068&
0.875&
 ...&
 ...&
 ...&
 ...\\
&4.60&
2.379&
0.233&
0.386&
4.640&
 ...&
 ...&
 ...\\
&4.70&
2.906&
0.540&
0.149&
3.042&
 ...&
 ...&
 ...\\
&4.80&
3.499&
0.917&
0.063&
1.850&
4.308&
 ...&
 ...\\
&4.90&
4.130&
1.335&
0.076&
0.965&
2.243&
 ...&
 ...\\
&5.00&
 ...&
1.932&
0.315&
0.468&
0.786&
 ...&
 ...\\
&5.10&
 ...&
3.025&
1.087&
0.633&
0.165&
 ...&
 ...\\
&5.20&
 ...&
4.344&
2.116&
1.152&
0.035&
 ...&
 ...\\
&5.30&
 ...&
 ...&
3.064&
1.664&
0.010&
 ...&
 ...\\
&5.40&
 ...&
 ...&
3.884&
2.107&
0.003&
 ...&
 ...\\
&5.50&
 ...&
 ...&
4.590&
2.483&
0.001&
 ...&
 ...\\
&5.60&
 ...&
 ...&
 ...&
2.794&
0.001&
3.929&
 ...\\
&5.70&
 ...&
 ...&
 ...&
3.023&
0.001&
2.761&
 ...\\
&5.80&
 ...&
 ...&
 ...&
3.148&
0.007&
1.803&
4.081& \\
&5.90&
 ...&
 ...&
 ...&
3.210&
0.044&
1.039&
2.395& \\
&6.00&
 ...&
 ...&
 ...&
3.353&
0.202&
0.528&
1.127& \\
&6.10&
 ...&
 ...&
 ...&
3.797&
0.658&
0.422&
0.396& \\
&6.20&
 ...&
 ...&
 ...&
4.534&
1.386&
0.674&
0.127& \\
&6.30&
 ...&
 ...&
 ...&
 ...&
2.150&
1.030&
0.046& \\
&6.40&
 ...&
 ...&
 ...&
 ...&
2.846&
1.374&
0.019& \\
&6.50&
 ...&
 ...&
 ...&
 ...&
3.463&
1.682&
0.009& \\
&6.60&
 ...&
 ...&
 ...&
 ...&
4.007&
1.952&
0.005& \\
&6.70&
 ...&
 ...&
 ...&
 ...&
4.493&
2.191&
0.003& \\
&6.80&
 ...&
 ...&
 ...&
 ...&
 ...&
2.403&
0.002& \\
&6.90&
 ...&
 ...&
 ...&
 ...&
 ...&
2.593&
0.001& \\
&7.00&
 ...&
 ...&
 ...&
 ...&
 ...&
2.766&
0.001& \\
&7.10&
 ...&
 ...&
 ...&
 ...&
 ...&
2.924&
0.001& \\
     \hline
 \end{tabular}
 }
 \newpage
 \centerline{Table 2. -- (continued) }
  
 {\tiny
 \begin{tabular}{ c c c c c c c c c c c}                                        
 \hline \hline
 &$\log_{10}$(T)&
 N I     &
 N II    &
 N III   &
 N IV    &
 N V     &
 N VI    &
 N VII   &
 N VIII  & \\
 \hline
&4.00&
0.001&
2.807&
 ...&
 ...&
 ...&
 ...&
 ...&
 ...\\
&4.10&
0.026&
1.231&
 ...&
 ...&
 ...&
 ...&
 ...&
 ...\\
&4.20&
0.344&
0.262&
 ...&
 ...&
 ...&
 ...&
 ...&
 ...\\
&4.30&
1.129&
0.034&
4.257&
 ...&
 ...&
 ...&
 ...&
 ...\\
&4.40&
1.787&
0.008&
2.717&
 ...&
 ...&
 ...&
 ...&
 ...\\
&4.50&
2.100&
0.014&
1.617&
 ...&
 ...&
 ...&
 ...&
 ...\\
&4.60&
2.279&
0.068&
0.854&
4.318&
 ...&
 ...&
 ...&
 ...\\
\hline
 \end{tabular}
 }
 \hspace{0.5cm}
  
 {\tiny
 \begin{tabular}{ c c c c c c c c c c c c}                                      
 \hline \hline
 &$\log_{10}$(T)&
 O I     &
 O II    &
 O III   &
 O IV    &
 O V     &
 O VI    &
 O VII   &
 O VIII  &
 O IX    & \\
 \hline
&4.00&
0.001&
2.510&
 ...&
 ...&
 ...&
 ...&
 ...&
 ...&
 ...\\
&4.10&
0.043&
1.028&
 ...&
 ...&
 ...&
 ...&
 ...&
 ...&
 ...\\
&4.20&
0.436&
0.198&
 ...&
 ...&
 ...&
 ...&
 ...&
 ...&
 ...\\
&4.30&
1.189&
0.029&
 ...&
 ...&
 ...&
 ...&
 ...&
 ...&
 ...\\
&4.40&
1.723&
0.008&
3.627&
 ...&
 ...&
 ...&
 ...&
 ...&
 ...\\
&4.50&
2.004&
0.007&
2.279&
 ...&
 ...&
 ...&
 ...&
 ...&
 ...\\
&4.60&
2.210&
0.023&
1.352&
 ...&
 ...&
 ...&
 ...&
 ...&
 ...\\
\hline
 \end{tabular}
 }
 \newpage
 \centerline{Table 2. -- (continued) }
  
 {\tiny
 \begin{tabular}{ c c c c c c c c c c c c c}                                    
 \hline \hline
 &$\log_{10}$(T)&
 F I     &
 F II    &
 F III   &
 F IV    &
 F V     &
 F VI    &
 F VII   &
 F VIII  &
 F IX    &
 F X     & \\
 \hline
&4.00&
0.000&
4.496&
 ...&
 ...&
 ...&
 ...&
 ...&
 ...&
 ...&
 ...\\
&4.10&
0.001&
2.581&
 ...&
 ...&
 ...&
 ...&
 ...&
 ...&
 ...&
 ...\\
&4.20&
0.038&
1.074&
 ...&
 ...&
 ...&
 ...&
 ...&
 ...&
 ...&
 ...\\
&4.30&
0.420&
0.208&
 ...&
 ...&
 ...&
 ...&
 ...&
 ...&
 ...&
 ...\\
&4.40&
1.231&
0.027&
3.277&
 ...&
 ...&
 ...&
 ...&
 ...&
 ...&
 ...\\
&4.50&
1.932&
0.010&
1.916&
 ...&
 ...&
 ...&
 ...&
 ...&
 ...&
 ...\\
&4.60&
2.383&
0.046&
1.015&
 ...&
 ...&
 ...&
 ...&
 ...&
 ...&
 ...\\
\hline
 \end{tabular}
 }
 \hspace{0.5cm}
  
 {\tiny
 \begin{tabular}{ c c c c c c c c c c c c c c}                                  
 \hline \hline
 &$\log_{10}$(T)&
 NeI     &
 NeII    &
 NeIII   &
 NeIV    &
 NeV     &
 NeVI    &
 NeVII   &
 NeVIII  &
 NeIX    &
 NeX     &
 NeXI    & \\
 \hline
&4.00&
0.000&
 ...&
 ...&
 ...&
 ...&
 ...&
 ...&
 ...&
 ...&
 ...&
 ...\\
&4.10&
0.000&
4.502&
 ...&
 ...&
 ...&
 ...&
 ...&
 ...&
 ...&
 ...&
 ...\\
&4.20&
0.001&
2.606&
 ...&
 ...&
 ...&
 ...&
 ...&
 ...&
 ...&
 ...&
 ...\\
&4.30&
0.035&
1.106&
 ...&
 ...&
 ...&
 ...&
 ...&
 ...&
 ...&
 ...&
 ...\\
&4.40&
0.398&
0.222&
4.816&
 ...&
 ...&
 ...&
 ...&
 ...&
 ...&
 ...&
 ...\\
&4.50&
1.195&
0.029&
2.843&
 ...&
 ...&
 ...&
 ...&
 ...&
 ...&
 ...&
 ...\\
&4.60&
1.858&
0.021&
1.482&
 ...&
 ...&
 ...&
 ...&
 ...&
 ...&
 ...&
 ...\\
\hline
 \end{tabular}
 }
 \newpage
 \centerline{Table 2. -- (continued) }
  
 {\tiny
 \begin{tabular}{ c c c c c c c c c c c c c c c}                                
 \hline \hline
 &$\log_{10}$(T)&
 NaI     &
 NaII    &
 NaIII   &
 NaIV    &
 NaV     &
 NaVI    &
 NaVII   &
 NaVIII  &
 NaIX    &
 NaX     &
 NaXI    &
 NaXII   & \\
 \hline
&4.00&
2.997&
0.000&
 ...&
 ...&
 ...&
 ...&
 ...&
 ...&
 ...&
 ...&
 ...&
 ...\\
&4.10&
3.672&
0.000&
 ...&
 ...&
 ...&
 ...&
 ...&
 ...&
 ...&
 ...&
 ...&
 ...\\
&4.20&
4.237&
0.000&
 ...&
 ...&
 ...&
 ...&
 ...&
 ...&
 ...&
 ...&
 ...&
 ...\\
&4.30&
4.713&
0.000&
 ...&
 ...&
 ...&
 ...&
 ...&
 ...&
 ...&
 ...&
 ...&
 ...\\
&4.40&
 ...&
0.000&
 ...&
 ...&
 ...&
 ...&
 ...&
 ...&
 ...&
 ...&
 ...&
 ...\\
&4.50&
 ...&
0.000&
4.592&
 ...&
 ...&
 ...&
 ...&
 ...&
 ...&
 ...&
 ...&
 ...\\
&4.60&
 ...&
0.001&
2.784&
 ...&
 ...&
 ...&
 ...&
 ...&
 ...&
 ...&
 ...&
 ...\\
\hline
 \end{tabular}
 }
 \hspace{0.5cm}
  
 {\tiny
 \begin{tabular}{ c c c c c c c c c c c c c c c c}                              
 \hline \hline
 &$\log_{10}$(T)&
 MgI     &
 MgII    &
 MgIII   &
 MgIV    &
 MgV     &
 MgVI    &
 MgVII   &
 MgVIII  &
 MgIX    &
 MgX     &
 MgXI    &
 MgXII   &
 MgXIII  & \\
 \hline
&4.00&
0.835&
0.069&
3.755&
 ...&
 ...&
 ...&
 ...&
 ...&
 ...&
 ...&
 ...&
 ...&
 ...\\
&4.10&
1.330&
0.025&
2.032&
 ...&
 ...&
 ...&
 ...&
 ...&
 ...&
 ...&
 ...&
 ...&
 ...\\
&4.20&
1.857&
0.094&
0.744&
 ...&
 ...&
 ...&
 ...&
 ...&
 ...&
 ...&
 ...&
 ...&
 ...\\
&4.30&
2.743&
0.584&
0.132&
 ...&
 ...&
 ...&
 ...&
 ...&
 ...&
 ...&
 ...&
 ...&
 ...\\
&4.40&
3.875&
1.368&
0.019&
 ...&
 ...&
 ...&
 ...&
 ...&
 ...&
 ...&
 ...&
 ...&
 ...\\
&4.50&
 ...&
2.087&
0.004&
 ...&
 ...&
 ...&
 ...&
 ...&
 ...&
 ...&
 ...&
 ...&
 ...\\
&4.60&
 ...&
2.689&
0.001&
 ...&
 ...&
 ...&
 ...&
 ...&
 ...&
 ...&
 ...&
 ...&
 ...\\
\hline
 \end{tabular}
 }
 \newpage
 \centerline{Table 2. -- (continued) }
  
 {\tiny
 \begin{tabular}{ c c c c c c c c c c c c c c c c c}                            
 \hline \hline
 &$\log_{10}$(T)&
 AlI     &
 AlII    &
 AlIII   &
 AlIV    &
 AlV     &
 AlVI    &
 AlVII   &
 AlVIII  &
 AlIX    &
 AlX     &
 AlXI    &
 AlXII   &
 AlXIII  &
 AlXIV   & \\
 \hline
&4.00&
1.997&
0.004&
 ...&
 ...&
 ...&
 ...&
 ...&
 ...&
 ...&
 ...&
 ...&
 ...&
 ...&
 ...\\
&4.10&
2.183&
0.003&
4.078&
 ...&
 ...&
 ...&
 ...&
 ...&
 ...&
 ...&
 ...&
 ...&
 ...&
 ...\\
&4.20&
2.277&
0.003&
2.788&
 ...&
 ...&
 ...&
 ...&
 ...&
 ...&
 ...&
 ...&
 ...&
 ...&
 ...\\
&4.30&
2.378&
0.010&
1.745&
 ...&
 ...&
 ...&
 ...&
 ...&
 ...&
 ...&
 ...&
 ...&
 ...&
 ...\\
&4.40&
2.533&
0.056&
0.926&
3.393&
 ...&
 ...&
 ...&
 ...&
 ...&
 ...&
 ...&
 ...&
 ...&
 ...\\
&4.50&
2.846&
0.246&
0.394&
1.567&
 ...&
 ...&
 ...&
 ...&
 ...&
 ...&
 ...&
 ...&
 ...&
 ...\\
&4.60&
3.518&
0.782&
0.323&
0.445&
 ...&
 ...&
 ...&
 ...&
 ...&
 ...&
 ...&
 ...&
 ...&
 ...\\
\hline
 \end{tabular}
 }
 \hspace{0.5cm}
  
 {\tiny
 \begin{tabular}{ c c c c c c c c c c c c c c c c c c}                          
 \hline \hline
 &$\log_{10}$(T)&
 SiI     &
 SiII    &
 SiIII   &
 SiIV    &
 SiV     &
 SiVI    &
 SiVII   &
 SiVIII  &
 SiIX    &
 SiX     &
 SiXI    &
 SiXII   &
 SiXIII  &
 SiXIV   &
 SiXV    & \\
 \hline
&4.00&
1.024&
0.043&
4.192&
 ...&
 ...&
 ...&
 ...&
 ...&
 ...&
 ...&
 ...&
 ...&
 ...&
 ...&
 ...\\
&4.10&
1.829&
0.008&
2.539&
 ...&
 ...&
 ...&
 ...&
 ...&
 ...&
 ...&
 ...&
 ...&
 ...&
 ...&
 ...\\
&4.20&
2.463&
0.020&
1.391&
 ...&
 ...&
 ...&
 ...&
 ...&
 ...&
 ...&
 ...&
 ...&
 ...&
 ...&
 ...\\
&4.30&
2.951&
0.102&
0.680&
 ...&
 ...&
 ...&
 ...&
 ...&
 ...&
 ...&
 ...&
 ...&
 ...&
 ...&
 ...\\
&4.40&
3.400&
0.303&
0.299&
4.702&
 ...&
 ...&
 ...&
 ...&
 ...&
 ...&
 ...&
 ...&
 ...&
 ...&
 ...\\
&4.50&
3.866&
0.599&
0.126&
3.285&
 ...&
 ...&
 ...&
 ...&
 ...&
 ...&
 ...&
 ...&
 ...&
 ...&
 ...\\
&4.60&
4.339&
0.928&
0.058&
2.194&
 ...&
 ...&
 ...&
 ...&
 ...&
 ...&
 ...&
 ...&
 ...&
 ...&
 ...\\
\hline
 \end{tabular}
 }
 \newpage
 \centerline{Table 2. -- (continued) }
  
 {\tiny
 \begin{tabular}{ c c c c c c c c c c c}                                        
 \hline \hline
 &$\log_{10}$(T)&
 P I     &
 P II    &
 P III   &
 P IV    &
 P V     &
 P VI    &
 P VII   &
 P VIII  & \\
 \hline
&4.00&
0.217&
0.405&
 ...&
 ...&
 ...&
 ...&
 ...&
 ...\\
&4.10&
1.027&
0.043&
3.556&
 ...&
 ...&
 ...&
 ...&
 ...\\
&4.20&
1.910&
0.009&
2.064&
 ...&
 ...&
 ...&
 ...&
 ...\\
&4.30&
2.538&
0.033&
1.156&
 ...&
 ...&
 ...&
 ...&
 ...\\
&4.40&
2.906&
0.132&
0.583&
3.872&
 ...&
 ...&
 ...&
 ...\\
&4.50&
3.229&
0.352&
0.258&
2.579&
 ...&
 ...&
 ...&
 ...\\
&4.60&
3.618&
0.673&
0.117&
1.638&
 ...&
 ...&
 ...&
 ...\\
&4.70&
4.072&
1.053&
0.098&
0.946&
4.445&
 ...&
 ...&
 ...\\
&4.80&
4.610&
1.505&
0.199&
0.475&
2.929&
 ...&
 ...&
 ...\\
&4.90&
 ...&
2.050&
0.432&
0.218&
1.805&
3.189&
 ...&
 ...\\
&5.00&
 ...&
2.688&
0.790&
0.153&
1.016&
1.440&
 ...&
 ...\\
&5.10&
 ...&
3.626&
1.475&
0.467&
0.720&
0.362&
 ...&
 ...\\
&5.20&
 ...&
 ...&
2.653&
1.321&
1.051&
0.065&
4.285&
 ...\\
&5.30&
 ...&
 ...&
3.807&
2.184&
1.463&
0.019&
2.789&
 ...\\
&5.40&
 ...&
 ...&
4.722&
2.838&
1.717&
0.018&
1.677&
 ...\\
&5.50&
 ...&
 ...&
 ...&
3.361&
1.882&
0.071&
0.863&
3.294& \\
&5.60&
 ...&
 ...&
 ...&
3.950&
2.145&
0.265&
0.360&
1.886& \\
&5.70&
 ...&
 ...&
 ...&
4.717&
2.613&
0.657&
0.178&
0.956& \\
&5.80&
 ...&
 ...&
 ...&
 ...&
3.289&
1.236&
0.276&
0.436& \\
&5.90&
 ...&
 ...&
 ...&
 ...&
4.193&
2.022&
0.655&
0.299& \\
&6.00&
 ...&
 ...&
 ...&
 ...&
 ...&
3.002&
1.287&
0.497& \\
&6.10&
 ...&
 ...&
 ...&
 ...&
 ...&
4.125&
2.106&
0.946& \\
&6.20&
 ...&
 ...&
 ...&
 ...&
 ...&
 ...&
3.149&
1.669& \\
&6.30&
 ...&
 ...&
 ...&
 ...&
 ...&
 ...&
4.589&
2.827& \\
&6.40&
 ...&
 ...&
 ...&
 ...&
 ...&
 ...&
 ...&
4.429& \\
&6.50&
 ...&
 ...&
 ...&
 ...&
 ...&
 ...&
 ...&
 ...\\
     \hline
 \end{tabular}
 }
  
 {\tiny
 \begin{tabular}{ c c c c c c c c c c c}                                        
 \hline \hline
 &$\log_{10}$(T)&
 P IX    &
 P X     &
 P XI    &
 P XII   &
 P XIII  &
 P XIV   &
 P XV    &
 P XVI   & \\
 \hline
&5.70&
2.606&
 ...&
 ...&
 ...&
 ...&
 ...&
 ...&
 ...\\
&5.80&
1.346&
3.372&
 ...&
 ...&
 ...&
 ...&
 ...&
 ...\\
&5.90&
0.595&
1.904&
3.766&
 ...&
 ...&
 ...&
 ...&
 ...\\
&6.00&
0.283&
0.998&
2.167&
3.701&
 ...&
 ...&
 ...&
 ...\\
&6.10&
0.306&
0.527&
1.111&
2.017&
3.255&
4.484&
 ...&
 ...\\
&6.20&
0.667&
0.472&
0.558&
0.934&
1.589&
2.201&
 ...&
 ...\\
&6.30&
1.515&
0.964&
0.621&
0.545&
0.728&
0.830&
 ...&
 ...\\
\hline
 \end{tabular}
 }
 \newpage
 \centerline{Table 2. -- (continued) }
  
 {\tiny
 \begin{tabular}{ c c c c c c c c c c c c}                                      
 \hline \hline
 &$\log_{10}$(T)&
 S I     &
 S II    &
 S III   &
 S IV    &
 S V     &
 S VI    &
 S VII   &
 S VIII  &
 S IX    & \\
 \hline
&4.00&
0.158&
0.515&
 ...&
 ...&
 ...&
 ...&
 ...&
 ...&
 ...\\
&4.10&
0.881&
0.061&
 ...&
 ...&
 ...&
 ...&
 ...&
 ...&
 ...\\
&4.20&
1.765&
0.008&
3.139&
 ...&
 ...&
 ...&
 ...&
 ...&
 ...\\
&4.30&
2.430&
0.011&
1.663&
 ...&
 ...&
 ...&
 ...&
 ...&
 ...\\
&4.40&
2.846&
0.102&
0.682&
4.530&
 ...&
 ...&
 ...&
 ...&
 ...\\
&4.50&
3.256&
0.385&
0.233&
2.799&
 ...&
 ...&
 ...&
 ...&
 ...\\
&4.60&
3.711&
0.755&
0.094&
1.739&
 ...&
 ...&
 ...&
 ...&
 ...\\
&4.70&
4.180&
1.130&
0.081&
1.017&
3.880&
 ...&
 ...&
 ...&
 ...\\
&4.80&
4.706&
1.543&
0.173&
0.529&
2.495&
 ...&
 ...&
 ...&
 ...\\
&4.90&
 ...&
2.041&
0.391&
0.259&
1.475&
4.810&
 ...&
 ...&
 ...\\
&5.00&
 ...&
2.641&
0.741&
0.189&
0.773&
3.128&
 ...&
 ...&
 ...\\
&5.10&
 ...&
3.357&
1.229&
0.312&
0.358&
1.899&
2.894&
 ...&
 ...\\
&5.20&
 ...&
4.209&
1.869&
0.630&
0.216&
1.073&
1.233&
 ...&
 ...\\
&5.30&
 ...&
 ...&
2.874&
1.349&
0.536&
0.814&
0.293&
 ...&
 ...\\
&5.40&
 ...&
 ...&
4.278&
2.493&
1.333&
1.110&
0.059&
3.551&
 ...\\
&5.50&
 ...&
 ...&
 ...&
3.563&
2.094&
1.431&
0.022&
2.367&
 ...\\
&5.60&
 ...&
 ...&
 ...&
4.437&
2.690&
1.635&
0.028&
1.435&
3.994& \\
&5.70&
 ...&
 ...&
 ...&
 ...&
3.229&
1.817&
0.100&
0.732&
2.369& \\
&5.80&
 ...&
 ...&
 ...&
 ...&
3.882&
2.145&
0.340&
0.327&
1.200& \\
&5.90&
 ...&
 ...&
 ...&
 ...&
4.775&
2.740&
0.841&
0.287&
0.523& \\
&6.00&
 ...&
 ...&
 ...&
 ...&
 ...&
3.634&
1.627&
0.618&
0.321& \\
&6.10&
 ...&
 ...&
 ...&
 ...&
 ...&
4.775&
2.647&
1.251&
0.505& \\
&6.20&
 ...&
 ...&
 ...&
 ...&
 ...&
 ...&
3.807&
2.078&
0.951& \\
&6.30&
 ...&
 ...&
 ...&
 ...&
 ...&
 ...&
 ...&
3.091&
1.635& \\
&6.40&
 ...&
 ...&
 ...&
 ...&
 ...&
 ...&
 ...&
4.432&
2.690& \\
&6.50&
 ...&
 ...&
 ...&
 ...&
 ...&
 ...&
 ...&
 ...&
4.158& \\
     \hline
 \end{tabular}
 }
  
 {\tiny
 \begin{tabular}{ c c c c c c c c c c c}                                        
 \hline \hline
 &$\log_{10}$(T)&
 S X     &
 S XI    &
 S XII   &
 S XIII  &
 S XIV   &
 S XV    &
 S XVI   &
 S XVII  & \\
 \hline
&5.70&
4.900&
 ...&
 ...&
 ...&
 ...&
 ...&
 ...&
 ...\\
&5.80&
2.844&
 ...&
 ...&
 ...&
 ...&
 ...&
 ...&
 ...\\
&5.90&
1.430&
3.444&
 ...&
 ...&
 ...&
 ...&
 ...&
 ...\\
&6.00&
0.610&
1.946&
3.558&
 ...&
 ...&
 ...&
 ...&
 ...\\
&6.10&
0.276&
1.044&
2.074&
3.488&
 ...&
 ...&
 ...&
 ...\\
&6.20&
0.281&
0.571&
1.125&
1.954&
3.084&
4.198&
 ...&
 ...\\
&6.30&
0.589&
0.470&
0.625&
0.968&
1.564&
2.103&
 ...&
 ...\\
\hline
 \end{tabular}
 }
 \newpage
 \centerline{Table 2. -- (continued) }
  
 {\tiny
 \begin{tabular}{ c c c c c c c c c c c c}                                      
 \hline \hline
 &$\log_{10}$(T)&
 ClI     &
 ClII    &
 ClIII   &
 ClIV    &
 ClV     &
 ClVI    &
 ClVII   &
 ClVIII  &
 ClIX    & \\
 \hline
&4.00&
0.015&
1.457&
 ...&
 ...&
 ...&
 ...&
 ...&
 ...&
 ...\\
&4.10&
0.304&
0.298&
 ...&
 ...&
 ...&
 ...&
 ...&
 ...&
 ...\\
&4.20&
1.203&
0.028&
3.179&
 ...&
 ...&
 ...&
 ...&
 ...&
 ...\\
&4.30&
2.131&
0.015&
1.576&
 ...&
 ...&
 ...&
 ...&
 ...&
 ...\\
&4.40&
2.978&
0.128&
0.595&
4.614&
 ...&
 ...&
 ...&
 ...&
 ...\\
&4.50&
3.767&
0.439&
0.198&
2.612&
 ...&
 ...&
 ...&
 ...&
 ...\\
&4.60&
4.370&
0.841&
0.085&
1.471&
 ...&
 ...&
 ...&
 ...&
 ...\\
&4.70&
4.862&
1.276&
0.100&
0.816&
4.092&
 ...&
 ...&
 ...&
 ...\\
&4.80&
 ...&
1.757&
0.220&
0.422&
2.811&
 ...&
 ...&
 ...&
 ...\\
&4.90&
 ...&
2.294&
0.438&
0.210&
1.859&
4.809&
 ...&
 ...&
 ...\\
&5.00&
 ...&
2.873&
0.729&
0.130&
1.155&
3.244&
 ...&
 ...&
 ...\\
&5.10&
 ...&
3.496&
1.088&
0.162&
0.657&
2.025&
4.715&
 ...&
 ...\\
&5.20&
 ...&
4.197&
1.546&
0.329&
0.370&
1.129&
2.949&
4.299&
 ...\\
&5.30&
 ...&
 ...&
2.153&
0.673&
0.321&
0.563&
1.655&
2.135&
 ...\\
&5.40&
 ...&
 ...&
3.046&
1.327&
0.631&
0.430&
0.905&
0.654&
 ...\\
&5.50&
 ...&
 ...&
4.567&
2.627&
1.626&
1.040&
0.983&
0.109&
3.816& \\
&5.60&
 ...&
 ...&
 ...&
4.145&
2.869&
1.946&
1.422&
0.024&
2.571& \\
&5.70&
 ...&
 ...&
 ...&
 ...&
4.000&
2.774&
1.833&
0.018&
1.612& \\
&5.80&
 ...&
 ...&
 ...&
 ...&
 ...&
3.518&
2.199&
0.067&
0.872& \\
&5.90&
 ...&
 ...&
 ...&
 ...&
 ...&
4.294&
2.633&
0.256&
0.406& \\
&6.00&
 ...&
 ...&
 ...&
 ...&
 ...&
 ...&
3.279&
0.700&
0.302& \\
&6.10&
 ...&
 ...&
 ...&
 ...&
 ...&
 ...&
4.204&
1.447&
0.592& \\
&6.20&
 ...&
 ...&
 ...&
 ...&
 ...&
 ...&
 ...&
2.439&
1.197& \\
&6.30&
 ...&
 ...&
 ...&
 ...&
 ...&
 ...&
 ...&
3.571&
1.998& \\
&6.40&
 ...&
 ...&
 ...&
 ...&
 ...&
 ...&
 ...&
4.873&
3.011& \\
&6.50&
 ...&
 ...&
 ...&
 ...&
 ...&
 ...&
 ...&
 ...&
4.407& \\
     \hline
 \end{tabular}
 }
  
 {\tiny
 \begin{tabular}{ c c c c c c c c c c c c}                                      
 \hline \hline
 &$\log_{10}$(T)&
 ClX     &
 ClXI    &
 ClXII   &
 ClXIII  &
 ClXIV   &
 ClXV    &
 ClXVI   &
 ClXVII  &
 ClXVIII & \\
 \hline
&5.70&
4.181&
 ...&
 ...&
 ...&
 ...&
 ...&
 ...&
 ...&
 ...\\
&5.80&
2.531&
 ...&
 ...&
 ...&
 ...&
 ...&
 ...&
 ...&
 ...\\
&5.90&
1.309&
2.896&
 ...&
 ...&
 ...&
 ...&
 ...&
 ...&
 ...\\
&6.00&
0.577&
1.451&
3.366&
 ...&
 ...&
 ...&
 ...&
 ...&
 ...\\
&6.10&
0.340&
0.623&
1.883&
3.562&
 ...&
 ...&
 ...&
 ...&
 ...\\
&6.20&
0.502&
0.292&
1.005&
2.073&
3.365&
 ...&
 ...&
 ...&
 ...\\
&6.30&
0.928&
0.302&
0.553&
1.109&
1.852&
2.812&
3.793&
 ...&
 ...\\
\hline
 \end{tabular}
 }
 \newpage
 \centerline{Table 2. -- (continued) }
  
 {\tiny
 \begin{tabular}{ c c c c c c c c c c c c c}                                    
 \hline \hline
 &$\log_{10}$(T)&
 ArI     &
 ArII    &
 ArIII   &
 ArIV    &
 ArV     &
 ArVI    &
 ArVII   &
 ArVIII  &
 ArIX    &
 ArX     & \\
 \hline
&4.00&
0.000&
3.318&
 ...&
 ...&
 ...&
 ...&
 ...&
 ...&
 ...&
 ...\\
&4.10&
0.012&
1.570&
 ...&
 ...&
 ...&
 ...&
 ...&
 ...&
 ...&
 ...\\
&4.20&
0.239&
0.373&
4.734&
 ...&
 ...&
 ...&
 ...&
 ...&
 ...&
 ...\\
&4.30&
1.068&
0.040&
2.487&
 ...&
 ...&
 ...&
 ...&
 ...&
 ...&
 ...\\
&4.40&
2.031&
0.054&
0.969&
 ...&
 ...&
 ...&
 ...&
 ...&
 ...&
 ...\\
&4.50&
3.195&
0.444&
0.196&
2.577&
 ...&
 ...&
 ...&
 ...&
 ...&
 ...\\
&4.60&
4.383&
1.044&
0.065&
1.304&
 ...&
 ...&
 ...&
 ...&
 ...&
 ...\\
\hline
 \end{tabular}
 }
  
 {\tiny
 \begin{tabular}{ c c c c c c c c c c c c}                                      
 \hline \hline
 &$\log_{10}$(T)&
 ArXI    &
 ArXII   &
 ArXIII  &
 ArXIV   &
 ArXV    &
 ArXVI   &
 ArXVII  &
 ArXVIII &
 ArXIX   & \\
 \hline
&5.70&
 ...&
 ...&
 ...&
 ...&
 ...&
 ...&
 ...&
 ...&
 ...\\
&5.80&
3.675&
 ...&
 ...&
 ...&
 ...&
 ...&
 ...&
 ...&
 ...\\
&5.90&
2.364&
4.742&
 ...&
 ...&
 ...&
 ...&
 ...&
 ...&
 ...\\
&6.00&
1.408&
3.024&
 ...&
 ...&
 ...&
 ...&
 ...&
 ...&
 ...\\
&6.10&
0.761&
1.773&
3.472&
 ...&
 ...&
 ...&
 ...&
 ...&
 ...\\
&6.20&
0.426&
0.948&
2.059&
3.425&
 ...&
 ...&
 ...&
 ...&
 ...\\
&6.30&
0.396&
0.510&
1.142&
1.991&
3.122&
4.568&
 ...&
 ...&
 ...\\
\hline
 \end{tabular}
 }
 \newpage
 \centerline{Table 2. -- (continued) }
  
 {\tiny
 \begin{tabular}{ c c c c c c c c c c c c c}                                    
 \hline \hline
 &$\log_{10}$(T)&
 K I     &
 K II    &
 K III   &
 K IV    &
 K V     &
 K VI    &
 K VII   &
 K VIII  &
 K IX    &
 K X     & \\
 \hline
&4.00&
3.150&
0.000&
 ...&
 ...&
 ...&
 ...&
 ...&
 ...&
 ...&
 ...\\
&4.10&
3.717&
0.000&
 ...&
 ...&
 ...&
 ...&
 ...&
 ...&
 ...&
 ...\\
&4.20&
4.189&
0.000&
 ...&
 ...&
 ...&
 ...&
 ...&
 ...&
 ...&
 ...\\
&4.30&
4.585&
0.000&
3.881&
 ...&
 ...&
 ...&
 ...&
 ...&
 ...&
 ...\\
&4.40&
 ...&
0.003&
2.124&
 ...&
 ...&
 ...&
 ...&
 ...&
 ...&
 ...\\
&4.50&
 ...&
0.078&
0.783&
4.176&
 ...&
 ...&
 ...&
 ...&
 ...&
 ...\\
&4.60&
 ...&
0.536&
0.154&
2.115&
 ...&
 ...&
 ...&
 ...&
 ...&
 ...\\
\hline
 \end{tabular}
 }
  
 {\tiny
 \begin{tabular}{ c c c c c c c c c c c c c}                                    
 \hline \hline
 &$\log_{10}$(T)&
 K XI    &
 K XII   &
 K XIII  &
 K XIV   &
 K XV    &
 K XVI   &
 K XVII  &
 K XVIII &
 K XIX   &
 K XX    & \\
 \hline
&5.70&
3.572&
 ...&
 ...&
 ...&
 ...&
 ...&
 ...&
 ...&
 ...&
 ...\\
&5.80&
2.334&
 ...&
 ...&
 ...&
 ...&
 ...&
 ...&
 ...&
 ...&
 ...\\
&5.90&
1.585&
3.676&
 ...&
 ...&
 ...&
 ...&
 ...&
 ...&
 ...&
 ...\\
&6.00&
1.047&
2.441&
4.550&
 ...&
 ...&
 ...&
 ...&
 ...&
 ...&
 ...\\
&6.10&
0.668&
1.512&
2.848&
 ...&
 ...&
 ...&
 ...&
 ...&
 ...&
 ...\\
&6.20&
0.467&
0.860&
1.558&
3.084&
 ...&
 ...&
 ...&
 ...&
 ...&
 ...\\
&6.30&
0.528&
0.540&
0.711&
1.657&
2.991&
4.526&
 ...&
 ...&
 ...&
 ...\\
\hline
 \end{tabular}
 }
 \newpage
 \centerline{Table 2. -- (continued) }
  
 {\tiny
 \begin{tabular}{ c c c c c c c c c c c c c c}                                  
 \hline \hline
 &$\log_{10}$(T)&
 CaI     &
 CaII    &
 CaIII   &
 CaIV    &
 CaV     &
 CaVI    &
 CaVII   &
 CaVIII  &
 CaIX    &
 CaX     &
 CaXI    & \\
 \hline
&4.00&
1.112&
0.041&
1.914&
 ...&
 ...&
 ...&
 ...&
 ...&
 ...&
 ...&
 ...\\
&4.10&
1.707&
0.123&
0.642&
 ...&
 ...&
 ...&
 ...&
 ...&
 ...&
 ...&
 ...\\
&4.20&
2.711&
0.683&
0.102&
 ...&
 ...&
 ...&
 ...&
 ...&
 ...&
 ...&
 ...\\
&4.30&
3.912&
1.493&
0.014&
 ...&
 ...&
 ...&
 ...&
 ...&
 ...&
 ...&
 ...\\
&4.40&
 ...&
2.219&
0.003&
 ...&
 ...&
 ...&
 ...&
 ...&
 ...&
 ...&
 ...\\
&4.50&
 ...&
2.817&
0.001&
4.496&
 ...&
 ...&
 ...&
 ...&
 ...&
 ...&
 ...\\
&4.60&
 ...&
3.200&
0.001&
2.798&
 ...&
 ...&
 ...&
 ...&
 ...&
 ...&
 ...\\
\hline
 \end{tabular}
 }
  
 {\tiny
 \begin{tabular}{ c c c c c c c c c c c c c}                                    
 \hline \hline
 &$\log_{10}$(T)&
 CaXII   &
 CaXIII  &
 CaXIV   &
 CaXV    &
 CaXVI   &
 CaXVII  &
 CaXVIII &
 CaXIX   &
 CaXX    &
 CaXXI   & \\
 \hline
&5.70&
 ...&
 ...&
 ...&
 ...&
 ...&
 ...&
 ...&
 ...&
 ...&
 ...\\
&5.80&
3.691&
 ...&
 ...&
 ...&
 ...&
 ...&
 ...&
 ...&
 ...&
 ...\\
&5.90&
2.338&
 ...&
 ...&
 ...&
 ...&
 ...&
 ...&
 ...&
 ...&
 ...\\
&6.00&
1.606&
3.587&
 ...&
 ...&
 ...&
 ...&
 ...&
 ...&
 ...&
 ...\\
&6.10&
1.086&
2.420&
4.308&
 ...&
 ...&
 ...&
 ...&
 ...&
 ...&
 ...\\
&6.20&
0.711&
1.529&
2.698&
4.682&
 ...&
 ...&
 ...&
 ...&
 ...&
 ...\\
&6.30&
0.503&
0.893&
1.470&
2.801&
4.379&
 ...&
 ...&
 ...&
 ...&
 ...\\
\hline
 \end{tabular}
 }
 \newpage
 \centerline{Table 2. -- (continued) }
  
 {\tiny
 \begin{tabular}{ c c c c c c c c c c c c c c}                                  
 \hline \hline
 &$\log_{10}$(T)&
 ScI     &
 ScII    &
 ScIII   &
 ScIV    &
 ScV     &
 ScVI    &
 ScVII   &
 ScVIII  &
 ScIX    &
 ScX     &
 ScXI    & \\
 \hline
&4.00&
0.894&
0.060&
3.007&
 ...&
 ...&
 ...&
 ...&
 ...&
 ...&
 ...&
 ...\\
&4.10&
1.268&
0.033&
1.726&
 ...&
 ...&
 ...&
 ...&
 ...&
 ...&
 ...&
 ...\\
&4.20&
1.652&
0.053&
1.030&
 ...&
 ...&
 ...&
 ...&
 ...&
 ...&
 ...&
 ...\\
&4.30&
2.063&
0.134&
0.591&
3.375&
 ...&
 ...&
 ...&
 ...&
 ...&
 ...&
 ...\\
&4.40&
2.534&
0.301&
0.322&
1.697&
 ...&
 ...&
 ...&
 ...&
 ...&
 ...&
 ...\\
&4.50&
3.180&
0.667&
0.306&
0.538&
 ...&
 ...&
 ...&
 ...&
 ...&
 ...&
 ...\\
&4.60&
4.255&
1.482&
0.781&
0.096&
 ...&
 ...&
 ...&
 ...&
 ...&
 ...&
 ...\\
\hline
 \end{tabular}
 }
  
 {\tiny
 \begin{tabular}{ c c c c c c c c c c c c c c}                                  
 \hline \hline
 &$\log_{10}$(T)&
 ScXII   &
 ScXIII  &
 ScXIV   &
 ScXV    &
 ScXVI   &
 ScXVII  &
 ScXVIII &
 ScXIX   &
 ScXX    &
 ScXXI   &
 ScXXII  & \\
 \hline
&5.70&
4.517&
 ...&
 ...&
 ...&
 ...&
 ...&
 ...&
 ...&
 ...&
 ...&
 ...\\
&5.80&
2.138&
 ...&
 ...&
 ...&
 ...&
 ...&
 ...&
 ...&
 ...&
 ...&
 ...\\
&5.90&
0.600&
3.523&
 ...&
 ...&
 ...&
 ...&
 ...&
 ...&
 ...&
 ...&
 ...\\
&6.00&
0.134&
2.256&
4.858&
 ...&
 ...&
 ...&
 ...&
 ...&
 ...&
 ...&
 ...\\
&6.10&
0.067&
1.565&
3.414&
 ...&
 ...&
 ...&
 ...&
 ...&
 ...&
 ...&
 ...\\
&6.20&
0.075&
1.070&
2.325&
4.024&
 ...&
 ...&
 ...&
 ...&
 ...&
 ...&
 ...\\
&6.30&
0.137&
0.712&
1.484&
2.590&
4.337&
 ...&
 ...&
 ...&
 ...&
 ...&
 ...\\
\hline
 \end{tabular}
 }
 \newpage
 \centerline{Table 2. -- (continued) }
  
 {\tiny
 \begin{tabular}{ c c c c c c c c c c c c c c c}                                
 \hline \hline
 &$\log_{10}$(T)&
 TiI     &
 TiII    &
 TiIII   &
 TiIV    &
 TiV     &
 TiVI    &
 TiVII   &
 TiVIII  &
 TiIX    &
 TiX     &
 TiXI    &
 TiXII   & \\
 \hline
&4.00&
2.138&
0.003&
3.310&
 ...&
 ...&
 ...&
 ...&
 ...&
 ...&
 ...&
 ...&
 ...\\
&4.10&
2.464&
0.010&
1.739&
 ...&
 ...&
 ...&
 ...&
 ...&
 ...&
 ...&
 ...&
 ...\\
&4.20&
2.653&
0.136&
0.575&
 ...&
 ...&
 ...&
 ...&
 ...&
 ...&
 ...&
 ...&
 ...\\
&4.30&
3.250&
0.678&
0.103&
3.694&
 ...&
 ...&
 ...&
 ...&
 ...&
 ...&
 ...&
 ...\\
&4.40&
3.910&
1.253&
0.026&
2.508&
 ...&
 ...&
 ...&
 ...&
 ...&
 ...&
 ...&
 ...\\
&4.50&
4.333&
1.564&
0.023&
1.618&
 ...&
 ...&
 ...&
 ...&
 ...&
 ...&
 ...&
 ...\\
&4.60&
4.687&
1.785&
0.064&
0.924&
3.170&
 ...&
 ...&
 ...&
 ...&
 ...&
 ...&
 ...\\
\hline
 \end{tabular}
 }
  
 {\tiny
 \begin{tabular}{ c c c c c c c c c c c c c c}                                  
 \hline \hline
 &$\log_{10}$(T)&
 TiXIII  &
 TiXIV   &
 TiXV    &
 TiXVI   &
 TiXVII  &
 TiXVIII &
 TiXIX   &
 TiXX    &
 TiXXI   &
 TiXXII  &
 TiXXIII & \\
 \hline
&5.80&
4.217&
 ...&
 ...&
 ...&
 ...&
 ...&
 ...&
 ...&
 ...&
 ...&
 ...\\
&5.90&
2.007&
 ...&
 ...&
 ...&
 ...&
 ...&
 ...&
 ...&
 ...&
 ...&
 ...\\
&6.00&
0.672&
3.379&
 ...&
 ...&
 ...&
 ...&
 ...&
 ...&
 ...&
 ...&
 ...\\
&6.10&
0.217&
2.202&
4.567&
 ...&
 ...&
 ...&
 ...&
 ...&
 ...&
 ...&
 ...\\
&6.20&
0.115&
1.524&
3.207&
 ...&
 ...&
 ...&
 ...&
 ...&
 ...&
 ...&
 ...\\
&6.30&
0.106&
1.041&
2.178&
3.648&
 ...&
 ...&
 ...&
 ...&
 ...&
 ...&
 ...\\
&6.40&
0.163&
0.698&
1.385&
2.318&
3.820&
 ...&
 ...&
 ...&
 ...&
 ...&
 ...\\
\hline
 \end{tabular}
 }
 \newpage
 \centerline{Table 2. -- (continued) }
  
 {\tiny
 \begin{tabular}{ c c c c c c c c c c c c c c c}                                
 \hline \hline
 &$\log_{10}$(T)&
 V I     &
 V II    &
 V III   &
 V IV    &
 V V     &
 V VI    &
 V VII   &
 V VIII  &
 V IX    &
 V X     &
 V XI    &
 V XII   & \\
 \hline
&4.00&
2.305&
0.002&
3.327&
 ...&
 ...&
 ...&
 ...&
 ...&
 ...&
 ...&
 ...&
 ...\\
&4.10&
2.987&
0.010&
1.685&
 ...&
 ...&
 ...&
 ...&
 ...&
 ...&
 ...&
 ...&
 ...\\
&4.20&
3.304&
0.155&
0.522&
4.672&
 ...&
 ...&
 ...&
 ...&
 ...&
 ...&
 ...&
 ...\\
&4.30&
3.768&
0.665&
0.108&
2.565&
 ...&
 ...&
 ...&
 ...&
 ...&
 ...&
 ...&
 ...\\
&4.40&
4.208&
1.139&
0.066&
1.166&
 ...&
 ...&
 ...&
 ...&
 ...&
 ...&
 ...&
 ...\\
&4.50&
4.731&
1.652&
0.279&
0.345&
4.548&
 ...&
 ...&
 ...&
 ...&
 ...&
 ...&
 ...\\
&4.60&
 ...&
2.307&
0.703&
0.100&
2.647&
 ...&
 ...&
 ...&
 ...&
 ...&
 ...&
 ...\\
\hline
 \end{tabular}
 }
  
 {\tiny
 \begin{tabular}{ c c c c c c c c c c c c c c c}                                
 \hline \hline
 &$\log_{10}$(T)&
 V XIII  &
 V XIV   &
 V XV    &
 V XVI   &
 V XVII  &
 V XVIII &
 V XIX   &
 V XX    &
 V XXI   &
 V XXII  &
 V XXIII &
 V XXIV  & \\
 \hline
&5.90&
3.091&
3.597&
 ...&
 ...&
 ...&
 ...&
 ...&
 ...&
 ...&
 ...&
 ...&
 ...\\
&6.00&
1.483&
1.482&
4.822&
 ...&
 ...&
 ...&
 ...&
 ...&
 ...&
 ...&
 ...&
 ...\\
&6.10&
0.765&
0.438&
2.930&
 ...&
 ...&
 ...&
 ...&
 ...&
 ...&
 ...&
 ...&
 ...\\
&6.20&
0.716&
0.171&
2.002&
4.142&
 ...&
 ...&
 ...&
 ...&
 ...&
 ...&
 ...&
 ...\\
&6.30&
0.827&
0.110&
1.408&
2.931&
4.797&
 ...&
 ...&
 ...&
 ...&
 ...&
 ...&
 ...\\
&6.40&
0.987&
0.115&
0.969&
1.989&
3.255&
 ...&
 ...&
 ...&
 ...&
 ...&
 ...&
 ...\\
&6.50&
1.208&
0.186&
0.662&
1.261&
2.043&
3.330&
4.800&
 ...&
 ...&
 ...&
 ...&
 ...\\
\hline
 \end{tabular}
 }
 \newpage
 \centerline{Table 2. -- (continued) }
  
 {\tiny
 \begin{tabular}{ c c c c c c c c c c c c c c c c}                              
 \hline \hline
 &$\log_{10}$(T)&
 CrI     &
 CrII    &
 CrIII   &
 CrIV    &
 CrV     &
 CrVI    &
 CrVII   &
 CrVIII  &
 CrIX    &
 CrX     &
 CrXI    &
 CrXII   &
 CrXIII  & \\
 \hline
&4.00&
2.377&
0.002&
4.229&
 ...&
 ...&
 ...&
 ...&
 ...&
 ...&
 ...&
 ...&
 ...&
 ...\\
&4.10&
3.255&
0.002&
2.414&
 ...&
 ...&
 ...&
 ...&
 ...&
 ...&
 ...&
 ...&
 ...&
 ...\\
&4.20&
4.004&
0.035&
1.109&
 ...&
 ...&
 ...&
 ...&
 ...&
 ...&
 ...&
 ...&
 ...&
 ...\\
&4.30&
4.617&
0.200&
0.434&
4.744&
 ...&
 ...&
 ...&
 ...&
 ...&
 ...&
 ...&
 ...&
 ...\\
&4.40&
 ...&
0.483&
0.174&
2.759&
 ...&
 ...&
 ...&
 ...&
 ...&
 ...&
 ...&
 ...&
 ...\\
&4.50&
 ...&
0.824&
0.094&
1.352&
 ...&
 ...&
 ...&
 ...&
 ...&
 ...&
 ...&
 ...&
 ...\\
&4.60&
 ...&
1.267&
0.175&
0.557&
3.697&
 ...&
 ...&
 ...&
 ...&
 ...&
 ...&
 ...&
 ...\\
\hline
 \end{tabular}
 }
  
 {\tiny
 \begin{tabular}{ c c c c c c c c c c c c c c c}                                
 \hline \hline
 &$\log_{10}$(T)&
 CrXIV   &
 CrXV    &
 CrXVI   &
 CrXVII  &
 CrXVIII &
 CrXIX   &
 CrXX    &
 CrXXI   &
 CrXXII  &
 CrXXIII &
 CrXXIV  &
 CrXXV   & \\
 \hline
&5.90&
4.889&
 ...&
 ...&
 ...&
 ...&
 ...&
 ...&
 ...&
 ...&
 ...&
 ...&
 ...\\
&6.00&
2.670&
3.164&
 ...&
 ...&
 ...&
 ...&
 ...&
 ...&
 ...&
 ...&
 ...&
 ...\\
&6.10&
1.255&
1.357&
4.383&
 ...&
 ...&
 ...&
 ...&
 ...&
 ...&
 ...&
 ...&
 ...\\
&6.20&
0.651&
0.465&
2.735&
 ...&
 ...&
 ...&
 ...&
 ...&
 ...&
 ...&
 ...&
 ...\\
&6.30&
0.621&
0.203&
1.871&
3.781&
 ...&
 ...&
 ...&
 ...&
 ...&
 ...&
 ...&
 ...\\
&6.40&
0.751&
0.133&
1.307&
2.658&
4.242&
 ...&
 ...&
 ...&
 ...&
 ...&
 ...&
 ...\\
&6.50&
0.937&
0.138&
0.896&
1.785&
2.831&
4.406&
 ...&
 ...&
 ...&
 ...&
 ...&
 ...\\
\hline
 \end{tabular}
 }
 \newpage
 \centerline{Table 2. -- (continued) }
  
 {\tiny
 \begin{tabular}{ c c c c c c c c c c c c c c c c}                              
 \hline \hline
 &$\log_{10}$(T)&
 MnI     &
 MnII    &
 MnIII   &
 MnIV    &
 MnV     &
 MnVI    &
 MnVII   &
 MnVIII  &
 MnIX    &
 MnX     &
 MnXI    &
 MnXII   &
 MnXIII  & \\
 \hline
&4.00&
1.974&
0.005&
4.026&
 ...&
 ...&
 ...&
 ...&
 ...&
 ...&
 ...&
 ...&
 ...&
 ...\\
&4.10&
2.801&
0.003&
2.266&
 ...&
 ...&
 ...&
 ...&
 ...&
 ...&
 ...&
 ...&
 ...&
 ...\\
&4.20&
3.284&
0.059&
0.895&
 ...&
 ...&
 ...&
 ...&
 ...&
 ...&
 ...&
 ...&
 ...&
 ...\\
&4.30&
3.791&
0.473&
0.178&
 ...&
 ...&
 ...&
 ...&
 ...&
 ...&
 ...&
 ...&
 ...&
 ...\\
&4.40&
4.424&
1.094&
0.037&
3.180&
 ...&
 ...&
 ...&
 ...&
 ...&
 ...&
 ...&
 ...&
 ...\\
&4.50&
4.837&
1.484&
0.020&
1.886&
 ...&
 ...&
 ...&
 ...&
 ...&
 ...&
 ...&
 ...&
 ...\\
&4.60&
 ...&
1.759&
0.046&
1.085&
4.574&
 ...&
 ...&
 ...&
 ...&
 ...&
 ...&
 ...&
 ...\\
\hline
 \end{tabular}
 }
  
 {\tiny
 \begin{tabular}{ c c c c c c c c c c c c c c c c}                              
 \hline \hline
 &$\log_{10}$(T)&
 MnXIV   &
 MnXV    &
 MnXVI   &
 MnXVII  &
 MnXVIII &
 MnXIX   &
 MnXX    &
 MnXXI   &
 MnXXII  &
 MnXXIII &
 MnXXIV  &
 MnXXV   &
 MnXXVI  & \\
 \hline
&6.00&
2.841&
4.284&
 ...&
 ...&
 ...&
 ...&
 ...&
 ...&
 ...&
 ...&
 ...&
 ...&
 ...\\
&6.10&
1.399&
2.232&
2.711&
 ...&
 ...&
 ...&
 ...&
 ...&
 ...&
 ...&
 ...&
 ...&
 ...\\
&6.20&
0.682&
0.993&
1.145&
3.870&
 ...&
 ...&
 ...&
 ...&
 ...&
 ...&
 ...&
 ...&
 ...\\
&6.30&
0.684&
0.555&
0.446&
2.491&
4.791&
 ...&
 ...&
 ...&
 ...&
 ...&
 ...&
 ...&
 ...\\
&6.40&
1.056&
0.560&
0.226&
1.722&
3.400&
 ...&
 ...&
 ...&
 ...&
 ...&
 ...&
 ...&
 ...\\
&6.50&
1.495&
0.695&
0.158&
1.197&
2.367&
3.716&
 ...&
 ...&
 ...&
 ...&
 ...&
 ...&
 ...\\
&6.60&
1.949&
0.894&
0.169&
0.820&
1.565&
2.434&
3.757&
 ...&
 ...&
 ...&
 ...&
 ...&
 ...\\
\hline
 \end{tabular}
 }
 \newpage
 \centerline{Table 2. -- (continued) }
  
 {\tiny
 \begin{tabular}{ c c c c c c c c c c c c c c c c c}                            
 \hline \hline
 &$\log_{10}$(T)&
 FeI     &
 FeII    &
 FeIII   &
 FeIV    &
 FeV     &
 FeVI    &
 FeVII   &
 FeVIII  &
 FeIX    &
 FeX     &
 FeXI    &
 FeXII   &
 FeXIII  &
 FeXIV   & \\
 \hline
&4.00&
0.779&
0.079&
4.037&
 ...&
 ...&
 ...&
 ...&
 ...&
 ...&
 ...&
 ...&
 ...&
 ...&
 ...\\
&4.10&
1.264&
0.027&
2.179&
 ...&
 ...&
 ...&
 ...&
 ...&
 ...&
 ...&
 ...&
 ...&
 ...&
 ...\\
&4.20&
1.760&
0.087&
0.783&
 ...&
 ...&
 ...&
 ...&
 ...&
 ...&
 ...&
 ...&
 ...&
 ...&
 ...\\
&4.30&
2.638&
0.587&
0.131&
3.931&
 ...&
 ...&
 ...&
 ...&
 ...&
 ...&
 ...&
 ...&
 ...&
 ...\\
&4.40&
3.680&
1.293&
0.026&
2.124&
 ...&
 ...&
 ...&
 ...&
 ...&
 ...&
 ...&
 ...&
 ...&
 ...\\
&4.50&
4.498&
1.807&
0.079&
0.824&
 ...&
 ...&
 ...&
 ...&
 ...&
 ...&
 ...&
 ...&
 ...&
 ...\\
&4.60&
 ...&
2.438&
0.444&
0.196&
3.566&
 ...&
 ...&
 ...&
 ...&
 ...&
 ...&
 ...&
 ...&
 ...\\
\hline
 \end{tabular}
 }
  
 {\tiny
 \begin{tabular}{ c c c c c c c c c c c c c c c c}                              
 \hline \hline
 &$\log_{10}$(T)&
 FeXV    &
 FeXVI   &
 FeXVII  &
 FeXVIII &
 FeXIX   &
 FeXX    &
 FeXXI   &
 FeXXII  &
 FeXXIII &
 FeXXIV  &
 FeXXV   &
 FeXXVI  &
 FeXXVII & \\
 \hline
&6.00&
4.466&
 ...&
 ...&
 ...&
 ...&
 ...&
 ...&
 ...&
 ...&
 ...&
 ...&
 ...&
 ...\\
&6.10&
2.442&
3.654&
4.412&
 ...&
 ...&
 ...&
 ...&
 ...&
 ...&
 ...&
 ...&
 ...&
 ...\\
&6.20&
1.174&
1.837&
2.199&
 ...&
 ...&
 ...&
 ...&
 ...&
 ...&
 ...&
 ...&
 ...&
 ...\\
&6.30&
0.620&
0.827&
0.886&
3.317&
 ...&
 ...&
 ...&
 ...&
 ...&
 ...&
 ...&
 ...&
 ...\\
&6.40&
0.727&
0.556&
0.371&
2.177&
4.263&
 ...&
 ...&
 ...&
 ...&
 ...&
 ...&
 ...&
 ...\\
&6.50&
1.091&
0.604&
0.207&
1.508&
3.055&
4.734&
 ...&
 ...&
 ...&
 ...&
 ...&
 ...&
 ...\\
&6.60&
1.507&
0.751&
0.161&
1.042&
2.142&
3.271&
4.880&
 ...&
 ...&
 ...&
 ...&
 ...&
 ...\\
\hline
 \end{tabular}
 }
 \newpage
 \centerline{Table 2. -- (continued) }
  
 {\tiny
 \begin{tabular}{ c c c c c c c c c c c c c c c c c}                            
 \hline \hline
 &$\log_{10}$(T)&
 CoI     &
 CoII    &
 CoIII   &
 CoIV    &
 CoV     &
 CoVI    &
 CoVII   &
 CoVIII  &
 CoIX    &
 CoX     &
 CoXI    &
 CoXII   &
 CoXIII  &
 CoXIV   & \\
 \hline
&4.00&
0.410&
0.214&
3.969&
 ...&
 ...&
 ...&
 ...&
 ...&
 ...&
 ...&
 ...&
 ...&
 ...&
 ...\\
&4.10&
0.821&
0.077&
1.965&
 ...&
 ...&
 ...&
 ...&
 ...&
 ...&
 ...&
 ...&
 ...&
 ...&
 ...\\
&4.20&
1.323&
0.135&
0.657&
 ...&
 ...&
 ...&
 ...&
 ...&
 ...&
 ...&
 ...&
 ...&
 ...&
 ...\\
&4.30&
2.054&
0.485&
0.178&
4.488&
 ...&
 ...&
 ...&
 ...&
 ...&
 ...&
 ...&
 ...&
 ...&
 ...\\
&4.40&
2.800&
0.885&
0.063&
2.521&
 ...&
 ...&
 ...&
 ...&
 ...&
 ...&
 ...&
 ...&
 ...&
 ...\\
&4.50&
3.519&
1.282&
0.065&
1.064&
 ...&
 ...&
 ...&
 ...&
 ...&
 ...&
 ...&
 ...&
 ...&
 ...\\
&4.60&
4.428&
1.885&
0.331&
0.284&
2.889&
 ...&
 ...&
 ...&
 ...&
 ...&
 ...&
 ...&
 ...&
 ...\\
\hline
 \end{tabular}
 }
  
 {\tiny
 \begin{tabular}{ c c c c c c c c c c c c c c c c c}                            
 \hline \hline
 &$\log_{10}$(T)&
 CoXV    &
 CoXVI   &
 CoXVII  &
 CoXVIII &
 CoXIX   &
 CoXX    &
 CoXXI   &
 CoXXII  &
 CoXXIII &
 CoXXIV  &
 CoXXV   &
 CoXXVI  &
 CoXXVII &
 CoXXVIII& \\
 \hline
&6.10&
2.675&
3.978&
 ...&
 ...&
 ...&
 ...&
 ...&
 ...&
 ...&
 ...&
 ...&
 ...&
 ...&
 ...\\
&6.20&
1.315&
2.054&
2.922&
3.556&
 ...&
 ...&
 ...&
 ...&
 ...&
 ...&
 ...&
 ...&
 ...&
 ...\\
&6.30&
0.685&
0.932&
1.277&
1.567&
4.407&
 ...&
 ...&
 ...&
 ...&
 ...&
 ...&
 ...&
 ...&
 ...\\
&6.40&
0.844&
0.663&
0.577&
0.590&
2.752&
 ...&
 ...&
 ...&
 ...&
 ...&
 ...&
 ...&
 ...&
 ...\\
&6.50&
1.497&
0.942&
0.505&
0.285&
1.896&
3.672&
 ...&
 ...&
 ...&
 ...&
 ...&
 ...&
 ...&
 ...\\
&6.60&
2.216&
1.333&
0.611&
0.185&
1.338&
2.609&
3.879&
 ...&
 ...&
 ...&
 ...&
 ...&
 ...&
 ...\\
&6.70&
2.911&
1.741&
0.782&
0.168&
0.932&
1.778&
2.545&
3.861&
 ...&
 ...&
 ...&
 ...&
 ...&
 ...\\
\hline
 \end{tabular}
 }
 \newpage
 \centerline{Table 2. -- (continued) }
  
 {\tiny
 \begin{tabular}{ c c c c c c c c c c c c c c c c c c}                          
 \hline \hline
 &$\log_{10}$(T)&
 NiI     &
 NiII    &
 NiIII   &
 NiIV    &
 NiV     &
 NiVI    &
 NiVII   &
 NiVIII  &
 NiIX    &
 NiX     &
 NiXI    &
 NiXII   &
 NiXIII  &
 NiXIV   &
 NiXV    & \\
 \hline
&4.00&
0.790&
0.077&
4.755&
 ...&
 ...&
 ...&
 ...&
 ...&
 ...&
 ...&
 ...&
 ...&
 ...&
 ...&
 ...\\
&4.10&
1.615&
0.012&
2.696&
 ...&
 ...&
 ...&
 ...&
 ...&
 ...&
 ...&
 ...&
 ...&
 ...&
 ...&
 ...\\
&4.20&
2.323&
0.037&
1.116&
 ...&
 ...&
 ...&
 ...&
 ...&
 ...&
 ...&
 ...&
 ...&
 ...&
 ...&
 ...\\
&4.30&
3.094&
0.420&
0.209&
4.876&
 ...&
 ...&
 ...&
 ...&
 ...&
 ...&
 ...&
 ...&
 ...&
 ...&
 ...\\
&4.40&
4.007&
1.235&
0.027&
2.749&
 ...&
 ...&
 ...&
 ...&
 ...&
 ...&
 ...&
 ...&
 ...&
 ...&
 ...\\
&4.50&
4.626&
1.855&
0.035&
1.198&
 ...&
 ...&
 ...&
 ...&
 ...&
 ...&
 ...&
 ...&
 ...&
 ...&
 ...\\
&4.60&
 ...&
2.396&
0.319&
0.288&
3.027&
 ...&
 ...&
 ...&
 ...&
 ...&
 ...&
 ...&
 ...&
 ...&
 ...\\
\hline
 \end{tabular}
 }
  
 {\tiny
 \begin{tabular}{ c c c c c c c c c c c c c c c c c}                            
 \hline \hline
 &$\log_{10}$(T)&
 NiXVI   &
 NiXVII  &
 NiXVIII &
 NiXIX   &
 NiXX    &
 NiXXI   &
 NiXXII  &
 NiXXIII &
 NiXXIV  &
 NiXXV   &
 NiXXVI  &
 NiXXVII &
 NiXXVIII&
 NiXXIX  & \\
 \hline
&6.10&
4.644&
 ...&
 ...&
 ...&
 ...&
 ...&
 ...&
 ...&
 ...&
 ...&
 ...&
 ...&
 ...&
 ...\\
&6.20&
2.529&
3.608&
4.765&
 ...&
 ...&
 ...&
 ...&
 ...&
 ...&
 ...&
 ...&
 ...&
 ...&
 ...\\
&6.30&
1.176&
1.718&
2.310&
2.823&
 ...&
 ...&
 ...&
 ...&
 ...&
 ...&
 ...&
 ...&
 ...&
 ...\\
&6.40&
0.694&
0.770&
0.895&
1.106&
3.622&
 ...&
 ...&
 ...&
 ...&
 ...&
 ...&
 ...&
 ...&
 ...\\
&6.50&
1.070&
0.741&
0.489&
0.452&
2.363&
4.435&
 ...&
 ...&
 ...&
 ...&
 ...&
 ...&
 ...&
 ...\\
&6.60&
1.743&
1.062&
0.506&
0.254&
1.667&
3.191&
4.692&
 ...&
 ...&
 ...&
 ...&
 ...&
 ...&
 ...\\
&6.70&
2.425&
1.435&
0.630&
0.186&
1.178&
2.246&
3.198&
4.720&
 ...&
 ...&
 ...&
 ...&
 ...&
 ...\\
\hline
 \end{tabular}
 }